\documentclass[iop]{emulateapj}

\newcommand{\OI}{O\,{\sc i}}
\newcommand{\HI}{H\,{\sc i}}
\newcommand{\CI}{C\,{\sc i}}
\newcommand{\CII}{C\,{\sc ii}}
\newcommand{\HII}{H\,{\sc ii}}

\newcommand{\NIII}{N\,{\sc iii}}
\newcommand{\OIII}{O\,{\sc iii}}

\def\,{\thinspace}

\def\nh2{{\hbox {$n({\rm H_2})$}}}

\slugcomment{Accepted for publication in Astrophysical Journal. November 11, 2014}

\shorttitle{Spatial distribution of excited CO, H$_2$O, OH, O~and~C$^+$ in Orion BN/KL outflows}

\shortauthors{Goicoechea et al.}

\begin{document}

\title{\textit{HERSCHEL}\altaffilmark{*} FAR-INFRARED SPECTRAL-MAPPING OF ORION BN/KL OUTFLOWS:\\
SPATIAL DISTRIBUTION OF EXCITED CO, H$_2$O, OH, O~and~C$^+$ IN SHOCKED GAS}

\author{Javier R. Goicoechea\altaffilmark{1,2},
Luis Chavarr\'{\i}a\altaffilmark{2,3},  Jos\'e Cernicharo\altaffilmark{1,2}, David A. Neufeld\altaffilmark{4},\\ 
Roland Vavrek\altaffilmark{5}, Edwin A. Bergin\altaffilmark{6}, Sara Cuadrado\altaffilmark{1,2}, 
Pierre Encrenaz\altaffilmark{7},\\ Mireya Etxaluze\altaffilmark{1,2}, Gary J. Melnick\altaffilmark{8},
Edward Polehampton\altaffilmark{9,10}}
\email{jr.goicoechea@icmm.csic.es}

\altaffiltext{*}{\textit{Herschel} is an ESA space observatory with science instruments provided by European-led
Principal Investigator consortia and with important participation from NASA.}

\altaffiltext{1}{Instituto de Ciencia de Materiales de Madrid (ICMM-CSIC).
Sor Juana Ines de la Cruz 3, 28049 Cantoblanco, Madrid, Spain.}

\altaffiltext{2}{Centro de Astrobiolog\'{\i}a, CSIC-INTA, Ctra. de Torrej\'on a Ajalvir km 4,
E-28850 Madrid, Spain.}

\altaffiltext{3}{Universidad de Chile/CONICYT, Camino del Observatorio 1515, Las Condes, Santiago, Chile.}

\altaffiltext{4}{Department of Physics \& Astronomy, Johns Hopkins University 3400 
North Charles Street, Baltimore, MD 21218, USA}

\altaffiltext{5}{Herschel Science Center, ESA/ESAC, P.O. Box 78, Villanueva de la Ca\~nada, E-28691 Madrid, Spain}

\altaffiltext{6}{Department of Astronomy, University of Michigan, 500 Church Street, Ann Arbor, MI, 48109, USA}

\altaffiltext{7}{LERMA, UMR 8112 du CNRS, Observatoire de Paris, \'Ecole Normale Sup\'erieure, France}

\altaffiltext{8}{Harvard-Smithsonian Center for Astrophysics, 60 Garden Street, MS 66, Cambridge, MA 02138, USA}

\altaffiltext{9}{RAL Space, Rutherford Appleton Laboratory, Chilton, Didcot, Oxfordshire OX11 0QX, UK}

\altaffiltext{10}{Institute for Space Imaging Science, University of Lethbridge, 
4401 University Drive, Lethbridge, Alberta T1J 1B1, Canada}

\begin{abstract}
 We present $\sim$2$'$$\times$2$'$ spectral-maps of Orion~BN/KL outflows taken with \textit{Herschel} at $\sim$12$''$ resolution. 
For the first time in the  far-IR domain, we spatially resolve 
the emission associated with the bright H$_2$ shocked regions ``Peak~1'' and ``Peak~2'' from that of the Hot Core and  ambient cloud.
We analyze the $\sim$54-310\,$\mu$m spectra taken with the PACS and SPIRE spectrometers.
More than 100  lines are detected, most of them rotationally excited lines of $^{12}$CO (up to $J$=48-47), H$_2$O, OH, $^{13}$CO, and HCN.
Peaks 1/2 are characterized by a very high $L$(CO)/$L_{FIR}$$\approx$5$\times$10$^{-3}$ ratio
and a plethora of far-IR H$_2$O emission lines. The high-$J$ CO and OH lines are a factor $\approx$2 brighter toward Peak~1 whereas several
excited H$_2$O lines are $\lesssim$50$\%$  brighter
toward Peak~2.
%A simplified non-LTE  model  allowed us to constrain the dominant gas temperature components. %toward Peak~1.
 Most of the CO column density arises from $T_{\rm k}$$\sim$200-500\,K gas that we associate with low-velocity 
 shocks that fail to sputter grain ice mantles and show a maximum gas-phase  H$_2$O/CO$\lesssim$10$^{-2}$ 
abundance ratio. 
In addition, the very excited CO ($J$$>$35) and H$_2$O lines reveal a  hotter gas component ($T_{\rm k}$$\sim$2500\,K) from faster
(v$_{\rm S}$$>$25~\,km\,s$^{-1}$) shocks that are able to sputter the frozen-out H$_2$O and
 lead to  high  H$_2$O/CO$\gtrsim$1 abundance ratios. 
The H$_2$O and OH luminosities 
cannot be reproduced by shock models that assume high (undepleted) abundances of atomic oxygen in the preshock gas and/or neglect the presence of
UV radiation in the postshock gas. 
Although massive outflows are a common feature in other massive star-forming cores, Orion~BN/KL  seems more peculiar 
because of its higher molecular 
luminosities and strong  outflows caused by a recent explosive event.

\end{abstract}


\section{Introduction}

The Becklin-Neugebauer/Kleinmann-Low (BN/KL) region at the core of the Orion molecular cloud~1 (OMC1)
and behind the Orion Nebula stellar cluster \citep[][]{Gen89,Ode01} 
is the nearest ($\sim$414\,pc) and probably the most studied high-mass star-forming region  
\citep[][]{Men95, Rei07}.
In addition to source~BN, two more compact \HII\, sources separated by $\sim$3$''$ are known to exist at the
centre of Orion~BN/KL, 
sources $I$ and $n$. The 3 sources are within a region of $\sim$10$''$ ($\sim$0.02\,pc).
Their proper motions reveal that they run away from a common region, suggesting that BN, $I$, and $n$ were 
originally part of a multiple protostellar system that merged $\lesssim$1000~years ago  
\citep{Bal05,Gom05,Zap09,Bal11,Pen12a,Nis12}.
The consequences of such an explosive event are likely related to the high-velocity CO emission and the
bright,  wide-angle H$_2$ outflow observed in the region since the late 70's \citep[e.g.,][]{Kwa76,Bec78}.

The different physical conditions and velocity fields along the line of sight however, 
complicate the interpretation of Orion~BN/KL observations.
This is  especially true for the low angular resolution observations carried out with single-dish telescopes. 
For this reason, it is common to distinguish between different physical components in the
region \citep[see][]{Bla87,Ter10}: the ``ridge'' (extended quiescent molecular gas), the ``Hot Core'' (a collection of very dense and 
hot  clumps showing an extremely rich chemistry), 
and the ``plateau'' (a mixture of outflows, shocks, and interactions with the ambient cloud). 

Different supersonic molecular outflows  arise from the core of Orion~BN/KL. 
A low-velocity outflow expanding at $\sim$18\,km\,s$^{-1}$ and  showing
a northeast-southwest (NE-SW) orientation was  
inferred from the proper motions of H$_2$O masers \citep[e.g.,][]{Gen81} and is  thought to be driven by source~$I$ \citep{Beu08}.
More recent interferometric observations of the $^{13}$CO $J$=2-1 line suggest the presence of a roughly spherical
bubble expanding at $\sim$15\,km\,s$^{-1}$ and with its center of symmetry coinciding  with the 
region where the protostellar system presumably merged, the ``explosion center''  \citep{Zap11}.
On the other hand, an even more powerful, but less understood, wide-angle high-velocity outflow, expanding
at $\sim$30\,km\,s$^{-1}$ to
a hundred \,km\,s$^{-1}$ exists and shows a northwest-southeast (NW-SE) orientation
approximately perpendicular to the low-velocity outflow
\citep[e.g.,][]{All93,Ode01}.
The high-velocity outflow contains the two brightest infrared H$_2$ emission peaks in the sky:
Peak~1 and Peak~2 \citep{Bec78}, with a total H$_2$ line luminosity of 120$\pm$60\,L$\sun$  \citep{Ros00} and
extending over a $\sim$2$'$$\times$2$'$ area. 
The H$_2$~Peak~1 is located $\sim$30$''$ NW of the Hot Core, and is the
brightest H$_2$ lobe of the wide-angle outflow. {Peak~2 is $\sim$20$''$ SE of the Hot Core.
A detailed analysis of the  H$_2$ ro-vibrational line emission toward Peak~1 shows excitation 
temperatures increasing from $\sim$600\,K to
about 3200\,K \citep{Ros00}. Such extreme conditions are consistent with shock-excited material 
as the outflow(s) plunge into the ambient cloud, heating and compressing the molecular gas to high temperatures
and densities. 

Ground-based observations of the  high-speed CO $J$=2-1  bullets  \citep{Rod99,Zap09} 
and of the $J$=6-5 and \mbox{7-6} high-velocity  emission \citep[][]{Pen12a,Pen12b} 
correlate with the  H$_2$ fingers seen in the near-IR, 
thus suggesting a common origin.
Therefore, in addition to being and excellent laboratory for star formation theories, the
environment around Orion~BN/KL, the H$_2$~Peaks in particular, are among the closest and brightest regions
to study the properties and processing of interstellar material by shock waves.

Early models of interstellar shocks in molecular clouds already predicted that, depending
on their nature and on the strength and orientation of the magnetic field, 
they will radiate high luminosities of molecular line emission 
\citep[H$_2$, H$_2$O and CO in particular, see e.g.,][]{Hol79,Dra83,Kau96,Ber98}.
Unfortunately, while the H$_2$, CO and H$_2$O vibrational spectrum can be observed toward  
sources of relatively low extinction \citep[see e.g.,][for mid-IR observations toward Peak~1]{Ros00,Gon02}, 
the relevant H$_2$O, CO and OH rotational  lines appear
at far-IR wavelengths that can only be observed from airborne or space telescopes. 
In particular, the water abundance and its role in the cooling of the warm interstellar gas has been a pivotal problem in any
discussion about molecular shocks \citep[see][for a comprehensive review]{vD13}.

The \textit{Kuiper Airborne Observatory} ($KAO$) and the \textit{Infrared Space Observatory} ($ISO$) 
opened the far-IR sky to  spectroscopic observations. 
The [\OI] fine structure lines, and the high-$J$ CO, H$_2$O and OH rotational lines
were first detected toward Orion BN/KL and other star-forming regions revealing shock-excited
material in many interstellar environments \citep[e.g.,][]{Gen89,Sch89,vD04,Cer05}.
Unfortunately, the small aperture of $KAO$ and $ISO$ greatly limited the sensitivity and the spatial
resolution of those pioneering observations even toward nearby regions like Orion.
The much higher angular and spectral capabilities provided by  \textit{Herschel} 3.5\,m~space telescope \citep{Pil10}
allowed us to carry out a  broadband spectral-mapping of the $\sim$2$'$$\times$2$'$ region
around Orion~BN/KL. For the first time at these wavelengths, we can separate the emission from the
shock-excited H$_2$~Peaks  from that of the central Hot Core regions where most
(sub)mm molecular lines peak.
The Orion H$_2$~Peaks~1/2  are  unique templates of a shock-excited environment where
 a  significant fraction of the gas cooling is due to far-IR lines.  In this work, we 
spatially resolve the
emission from all relevant far-IR cooling lines with an unprecedented $\sim$12$''$ ($\simeq$ 5000\,AU) resolution.

The spectral-maps presented here were taken with the PACS spectrometer
\citep{Pog10} as a part of the HEXOS GT Key Program, \textit{Herschel observations of EXtra-Ordinary Sources}
\citep{Ber10}. They were complemented with SPIRE-FTS \citep{Gri10} observations toward Orion~BN/KL. 
The paper is organized as follows. In Section~2 we present the data set and the data reduction technique.
In Section~3 we show the spatial distribution of CO, H$_2$O, OH and the atomic fine structure lines in the region
and also describe the nearly complete $\sim$54-310\,$\mu$m spectrum toward H$_2$~Peak~1/2.
In Section~4 we analyze maps of different line surface brightness ratios, construct CO, H$_2$O and OH rotational diagrams
and present a simple non-LTE model of the far-IR line emission toward Peak~1. In Section~5 we examine the line cooling
in the region and in Section~6 we discuss our results
in the context of  shock models available in the literature. We finally compare our observations with
those toward more distant high-mass star-forming cores.

\begin{figure*}[ht]
\begin{center}
\includegraphics[angle=0,scale=.81]{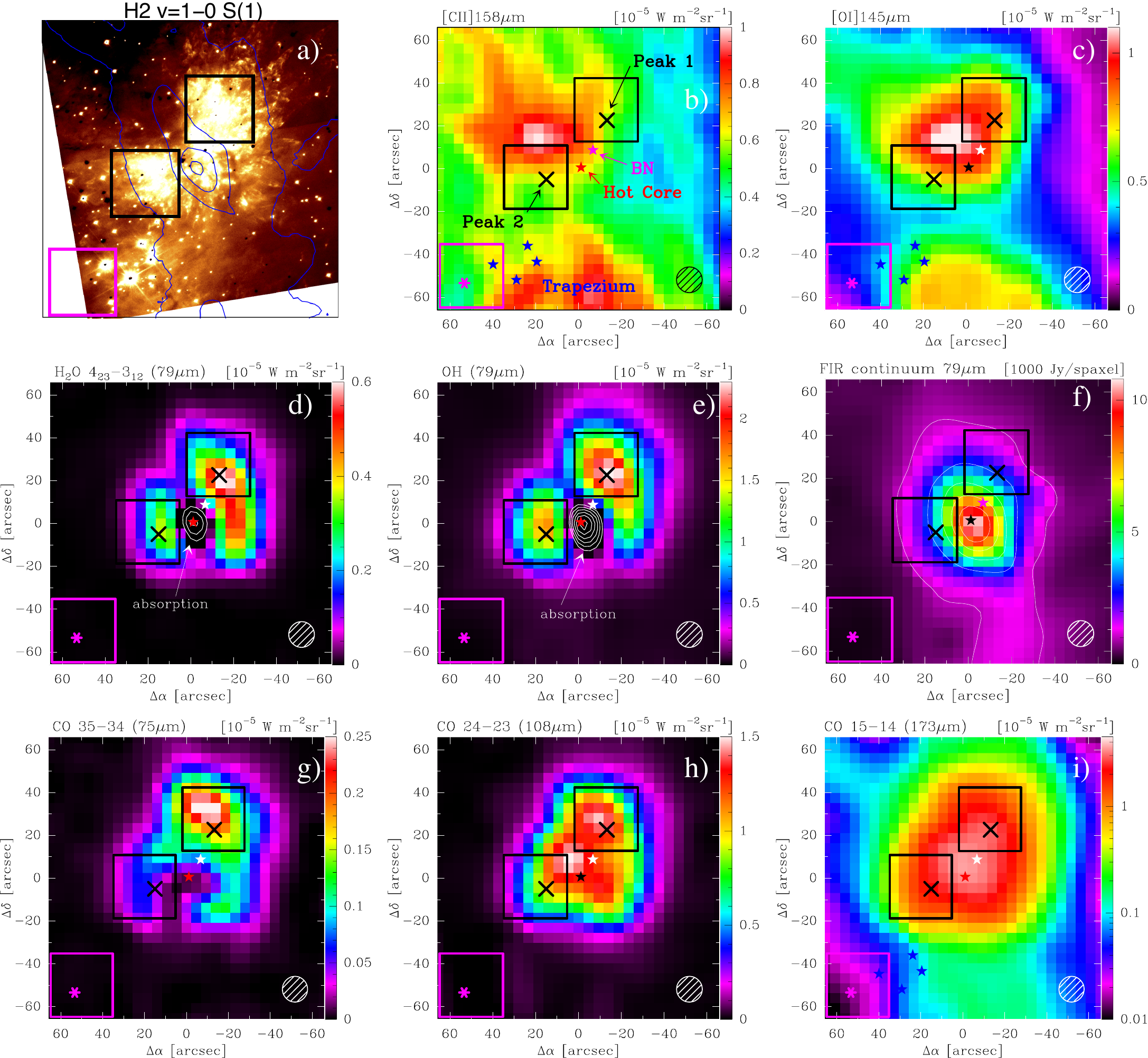}  %Apj
\hspace{0.5cm}
\caption{Line surface brightness maps obtained with \textit{Herschel}/PACS toward Orion BN/KL. 
The FoV is 2$'$$\times$2$'$.
The areas used to compute the line luminosities in the three representative regions discussed in the text  
(Peak~1, Peak~2 and ambient cloud)  are shown with black and pink squares respectively.
The black crosses and the pink star show the exact positions where the spectra shown in Figure~\ref{fig:pacs_spire_spectra} are extracted. 
The position of the BN source and that of the Orion Hot core are also shown in each panel as colored stars.
Panel $a)$ H$_{2}^{*}$ emission distribution in the region \citep{Bal11}. The 4 blue contours show the 
SCUBA 450\,$\mu$m continuum emission at 3$\sigma$ and 20, 50 and 90\% of the emission peak respectively \cite{Joh99}.
Panel $b)$ shows the [\CII]\,158$\mu$m line surface brightness. The position of the Trapezium stars are marked with blue stars.
The white contours in panels $d)$ and $e)$ represent H$_2$O and OH line absorption contours
toward the FIR continuum peak (from $-$0.1$\times$10$^{-5}$ to $-$0.4$\times$10$^{-5}$\,W\,m$^{-2}$\,sr$^{-1}$ in steps of 
0.1$\times$10$^{-5}$\,W\,m$^{-2}$\,sr$^{-1}$ for $o$-H$_2$O~4$_{23}$-3$_{12}$, and from 
$-$0.1$\times$10$^{-5}$ to $-$1.1$\times$10$^{-5}$\,W\,m$^{-2}$\,sr$^{-1}$ in steps of 
0.2$\times$10$^{-5}$\,W\,m$^{-2}$\,sr$^{-1}$ for OH $^2\Pi_{1/2}$-$^2\Pi_{3/2}$  $J$=1/2-3/2). Outside this region, lines are observed in emission.
Panel $f)$ shows the strong FIR continuum emission distribution at 79\,$\mu$m measured by the PACS spectrometer.
Panels $g)$, $h)$ and $i)$ show how the far-IR CO emission peak shifts toward Peak 1 
as the $J$  rotational number increases and the observed wavelength decreases.   
The $\sim$12$''$ angular resolution is shown in the bottom-right corner of each panel.} 
\label{fig:orion_maps_pacs}
\end{center}
\end{figure*}

%\clearpage

\section{Observations and Data Reduction}

\subsection{Non-standard PACS observations of BN/KL core}
\label{non-standard}

The PACS spectrometer was used to observe the core of Orion BN/KL 
at wavelengths between $\sim$54 and $\sim$190\,$\mu$m.
PACS used photoconductor detectors providing 25 spectra over a 47$''$$\times$47$''$ field of view (FoV) 
and resolved
5$\times$5 ``spaxels'', each with a size of 9.4$''$$\times$9.4$''$ in the sky.
The resolving power varies between $R$$\sim$1000-1500 ($\sim$108-190\,$\mu$m range),
$\sim$1700-3000 (70-94\,$\mu$m range) and $\sim$2700-5500 (54-70\,$\mu$m range).
The measured width of the PACS spectrometer point spread function (PSF)  is relatively constant for
$\lambda$$\lesssim$100\,$\mu$m ($\simeq$spaxel size) but increases at longer wavelengths.
In particular only between $\sim$74\%\, and 41\%\,of a point 
source emission would fall in a given spaxel between $\sim$54 and 190\,$\mu$m.

Owing to the very high far-IR continuum levels expected in the core of BN/KL (above 
the standard saturation limits of PACS) these observations were conducted in a
specific \textit{non-standard} engineering observing mode (\textit{PacsCalWaveCalNoChopBurst}).
In particular, the detector capacitances and the integration ramp lengths per grating scan were modified from their standard
values according to the expected  continuum fluxes.
Single-pointing observations with the PACS array were carried out in April 2011 using the ``unchopped'' mode.
The PACS array was centered at
$\alpha_{2000}$:~5$^h$35$^m$14.5$^s$, $\delta_{2000}$:~$-5^o$22$'$30.9$''$ in the Hot Core region
(ObsIDs 1342218575 and 1342218576). We checked that the observations were not affected by
the decrease of temperature of the star-tracker camera during orbits 320 and 761 and thus
the pointing is accurate within $\sim$2$''$.
Background subtraction was achieved 
by removing the telescope spectrum measured on a distant reference OFF-position separated by
$\sim$20$'$ in right ascension. 
Reference observations (ObsIDs 1342218573, 1342218574, 1342218577 and 1342218578)
were carried out before and after the ON-source observations.
The total observing time was 2.2\,h.
The resulting data were indeed above the standard PACS saturation limits, 
with continuum flux densities above 13000~Jy\,spaxel$^{-1}$ in the R1 grating order. The
response drifts due to self-curing were estimated to be within $\sim$15$\%$,
otherwise the data looked fine and  were validated for further scientific analysis.

\subsection{PACS 2$'$$\times$2$'$ spectral maps}

The PACS spectrometer was also used to map the $\sim$2$'$$\times$2$'$  region encompassing the Orion~BN/KL outflows
in the \mbox{70-94} and \mbox{108-190}\,$\mu$m  ranges (B2B and R1 grating orders respectively).
Nine individual pointings of the PACS array were done to cover the region in
a 3$\times$3 raster map with a step of one PACS FoV.
%The total number of observed positions in the map is 225.

Mapping observations were carried out in April 2011 using the range spectroscopy ``unchopped'' mode.
The map was also centered at the Hot Core position (ObsID 1342218572). 
Background subtraction was achieved 
by removing the telescope spectrum measured on the same reference OFF-position (ObsID 1342218571).
The reference OFF position spectrum showed a weak [\CII]\,158\,$\mu$m line emission contamination that was
removed before ON-OFF subtraction. 
Owing to the high far-IR continuum fluxes densities expected in the region (a few thousand Jy\,spaxel$^{-1}$), 
these mapping observations were conducted in the \textit{standard mode} but using the maximum capacitance allowed by the PACS detectors.
Still, the data below $\sim$130\,$\mu$m were saturated in the 25 spaxels of the central PACS array pointing.
These spectra were substituted by the \textit{non-standard} data discussed in Sect.~\ref{non-standard}.
The total mapping observing time was $\sim$3.4\,h.

\subsection{Line flux extraction}

PACS data were processed using version 8 of the Herschel Interactive Processing Environment (HIPE) 
and then exported to GILDAS where basic line spectrum manipulations were carried out.
The PACS flux calibration accuracy is limited by detector response drifts and slight pointing offsets.
The absolute flux calibration accuracy is determined by observations of 30 standard flux calibrators carried out
by the PACS instrument team and is estimated to be of the order of 
$\pm$30\%\footnote{\textit{PACS Spectroscopy performance and calibration}, PACS/ICC document ID PICC-KL-TN-041
(Vandenbussche et al.).}.
Additional processing of the spectral-maps was carried out using dedicated \textit{fortran90} subroutines.
In particular, the raster map spectra were gridded to a regular data cube through convolution
with a Gaussian kernel. The final angular resolution in the maps is $\sim$12$''$. Figure~\ref{fig:orion_maps_pacs} 
shows a selection of  PACS line and continuum maps and
Figure~\ref{fig:pacs_spire_spectra} shows high signal-to-noise (S/N) spectra  toward the
($-$13$''$,$+$23$''$) position in the Peak~1 region (black histograms) and toward the ($+$55$''$,$-$53$''$) position
in the ambient cloud (pink histograms in panels \ref{fig:pacs_spire_spectra}$a$ and \ref{fig:pacs_spire_spectra}$b$). 
Panels \ref{fig:pacs_spire_spectra}$c$, \ref{fig:pacs_spire_spectra}$d$ and \ref{fig:pacs_spire_spectra}$e$ 
show  the shortest-wavelength spectra
toward Peak~1 again  (black) and toward the ($+$19$''$,$-$4$''$) position in the Peak~2 region (gray).

Our PACS maps are not fully sampled spatially and the individual spaxels do not fill the PSF entirely.
For semi-extended sources this means that accurate line surface brightness can only be extracted
adding the fluxes measured in apertures that cover several spaxels.
The black and pink squares in Figure~\ref{fig:orion_maps_pacs} show the $\sim$30$''$$\times$30$''$ areas used to
extract the line fluxes between 70 and 190\,$\mu$m in the Peak~1/2 regions and in the ambient cloud. 
A third aperture around the (0$''$,0$''$) map position was defined to extract the CO line fluxes 
in the Hot Core region.
Only for lines  in the 54-70\,$\mu$m range (not mapped), 
line fluxes toward Peaks~1/2 were directly extracted from the ($-$13$''$,$+$23$''$) and  ($+$19$''$,$-$4$''$) 
spaxels ($\textit{non-standard}$ observations in Sect.~\ref{non-standard}) 
and then calibrated using the PACS point source correction. Owing to the almost constant PSF width at these short wavelengths,
the resulting line intensities are in good agreement (below the absolute
calibration accuracy) with those expected at longer wavelengths and extracted from the maps.  
Lines in the PACS wavelength  ranges affected by detector leakage were not included in the analysis.

\subsection{SPIRE FTS maps}

SPIRE FTS spectra between $\sim$194 and $\sim$671\,$\mu$m (1545-447\,GHz) 
were obtained on March 2011 in the high spectral 
resolution mode ($\Delta\tilde{\lambda}$=0.04\,cm$^{-1}$). The SPIRE FTS uses two bolometer arrays covering 
the 194-310\,$\mu$m  (Short Wavelength array, SSW) and 303-671\,$\mu$m  (Long Wavelength array, SLW) bands.
The two arrays contain 19 (SLW) and 37 (SSW) hexagonally packed detectors separated by 
$\sim$2 beams (51$"$ and 33$"$ respectively). The unvignetted FoV is $\sim$2$'$.
The SPIRE FTS observation (ObsID 1342216879) was centered at the same PACS map (0$''$,0$''$) position.
Observations were done in the \textit{bright source mode}.
The total observing time of the SPIRE sparse sampling map was 9\,min.

Owing to the extended nature of the line emission in Orion, we applied the \textit{extended emission}
pipeline to reduce the data with HIPE 11 \citep{Lu13}.  Once reduced and calibrated,
we used the unapodized spectra to fit the line intensities using sinc functions.
Opposite to PACS, the SPIRE pipeline provides the line surface brightnesses directly in W\,m$^{-2}$\,sr$^{-1}$. 
We checked that the  CO $J$=13-12 line intensities toward Peak~1, the Hot Core and the ambient cloud position
agree (within $\sim$30\%) with the line intensity expected from 
 the CO $J$=14-13 line intensities observed by PACS and extracted in $\sim$30$''$$\times$30$''$ apertures.
The 194-313\,$\mu$m spectra taken from the two SSW detectors
pointing toward  Peak~1 and the ambient postion are shown in Fig.~\ref{fig:pacs_spire_spectra}.
The units of Jy/beam in Fig.~\ref{fig:pacs_spire_spectra} were obtained by multiplying the spectrum 
in W\,m$^{-2}$\,Hz$^{-1}$\,sr$^{-1}$ by the effective beam size for extended sources \citep[see][]{Swi14}.
Note that no SLW  detector (303-671\,$\mu$m) matched the coordinates of the Peak~1 and ambient cloud positions 
observed with PACS thus they are not show in Figure~\ref{fig:pacs_spire_spectra}.
%They were just used to estimate the total line luminosities in the entire map.

\begin{figure*}[ht]
\begin{center} 
\includegraphics[angle=0,scale=1.12]{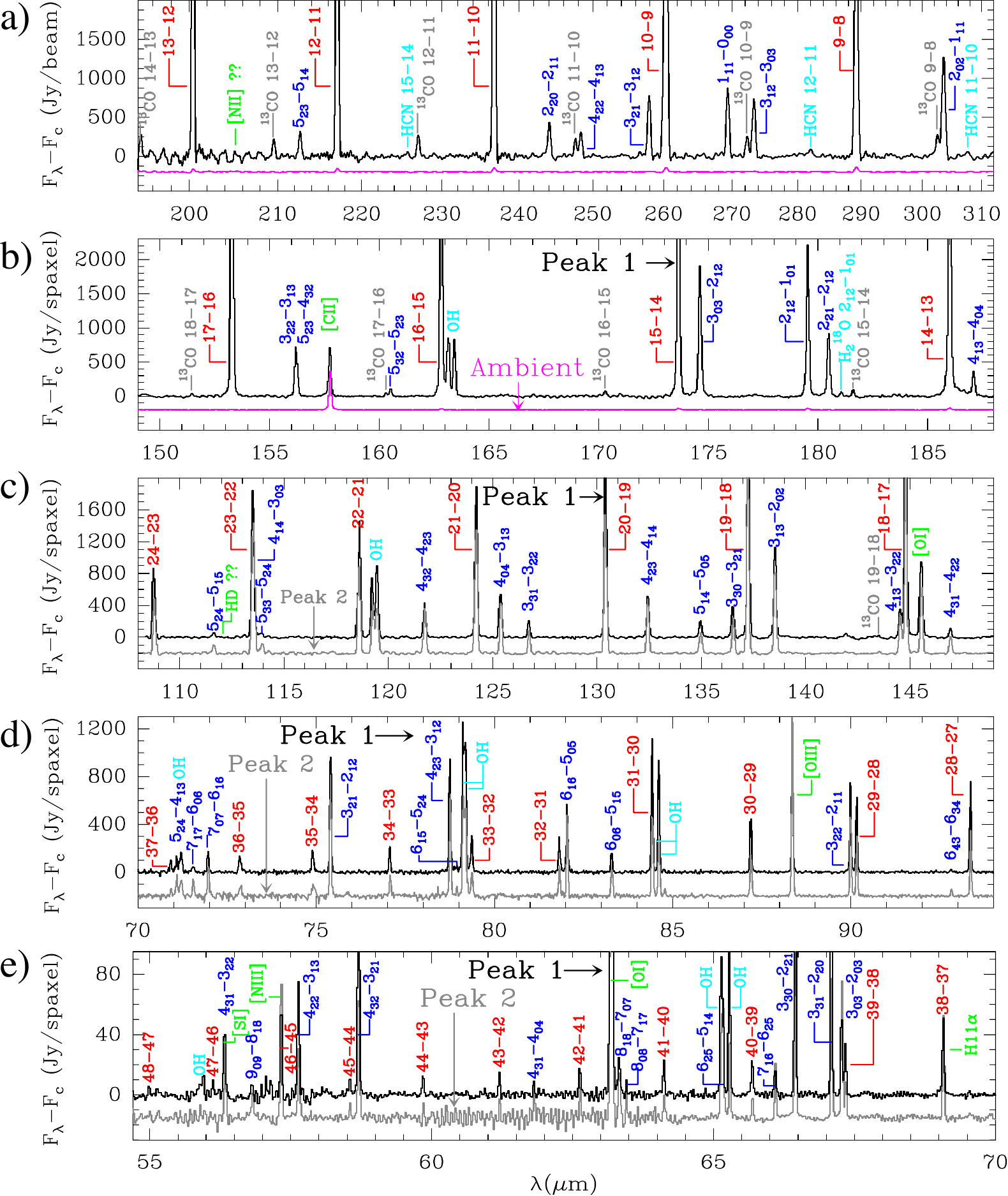} % ApJ
\hspace{0.0cm}
\caption{Panels $a)$ and $b)$: Continuum-subtracted SPIRE-FTS spectra (apodized for clarity in the figure)
and PACS spectra toward 
Orion H$_2$~Peak~1  (black) and toward an ambient cloud position (pink) (see Figure~\ref{fig:orion_maps_pacs}).
Panels $c)$, $d)$ and $e)$: Continuum-subtracted PACS spectra  toward 
Peak~1  (black)  and Peak~2 (gray). Note that the ambient cloud and Peak~2 spectra are shifted to ease the comparison.
Flux density units are in Jy/beam for SPIRE and  Jy/spaxel and for PACS (y-axis).
The x-axis shows the spectrum wavelength in microns. 
All detected spectral features are labelled ($^{12}$CO rotational lines in red, H$_2$O lines in blue, etc.).}
\label{fig:pacs_spire_spectra}
\end{center}
\end{figure*}

\clearpage

\section{Results}

\subsection{Spatial distribution of CO, H$_2$O, OH, O and C$^+$}

Figure~\ref{fig:orion_maps_pacs} shows a selection of $\sim$2$'$$\times$2$'$ continuum and line 
surface brightness maps taken with \textit{Herschel}/PACS. 
In these figures, the colored stars show the position of source BN and of the Hot Core (including the famous IRc2 source 
and radio sources $I$ and $n$). 
The crosses show the positions of the vibrationally excited H$_2$  emission peaks
in the region \citep[][]{Bec78}, referred as Peak~1 in the NW and Peak~2 in the SE 
(see Fig.\ref{fig:orion_maps_pacs}$a$).
The  far-IR dust continuum  peaks at the Hot Core position.
The strong continuum emitting region at $\sim$79\,$\mu$m has a radius of $\sim$15$''$
at half-maximum power  %(HMP).
and decreases outwards following the NE-SW direction along the ridge
(Fig.\ref{fig:orion_maps_pacs}$f$).

The spatial distribution of the CO line surface brightness is not homogeneous and depends on the $J$ rotational number.
The brightest $^{12}$CO lines are those between $J$=15 and 20. They show a 
roughly spherical distribution, with an approximated half-power radius
of $\sim$25$''$ (see Fig.~\ref{fig:orion_maps_pacs}$i$ for the $J$=15-14 line). 
The innermost regions of this spherical structure could correspond to the expanding bubble reported
by Zapata et al. (2011) 
and falling  at the center of the observed structure.
The $^{13}$CO $J$=16-15 and 15-14 lines are less affected by  opacity effects and they peak
slightly NE of the Hot Core. 
As $J$ increases, the CO spatial distribution changes. Figure~\ref{fig:orion_maps_pacs}$h$ shows
that the \mbox{$J$=24-23} line ($E_{\rm u}/k$=1658~K) peaks toward two different positions with similar integrated intensities,
one $\sim$10$''$ NE of the Hot Core (characteristic shocks produced by the low-velocity outflows) and  
other $\sim$30$''$ NW  in the H$_2$ Peak~1 region.  Peak~1 becomes much more apparent in the very high-$J$ line emission
maps (see  Fig.~\ref{fig:orion_maps_pacs}$g$ for the $J$=35-34 line with $E_{\rm u}/k$=3474~K).
Unlike other molecules emitting in the (sub)mm, the very high-$J$  $^{12}$CO line surface brightness maps 
resemble the vibrationally excited H$_{2}$ images, suggesting that they trace the same high-excitation shocked gas.
In fact, the very high-$J$ CO emission morphology revealed by PACS is similar to the high-velocity 
($|v_{lsr}|$$>$70\,km\,s$^{-1}$)
CO $J$=\mbox{7-6} and \mbox{6-5} line emission mapped with APEX \citep{Pen12a}
and reasonably follows the spatial distribution of the high-velocity  CO $J$=2-1 bullets
resolved by the SMA  interferometer. Interestingly, their distribution also show an emission ``hole'' inside the explosion center region
\citep{Zap09}. All together, this suggests that the much higher-energy far-IR CO lines are dominated 
by material moving at high speeds.

Most of the far-IR H$_2$O and OH  rotational lines are  sensitive to the strength of the 
dust continuum field. The angular resolution of our \textit{Herschel}/PACS maps is comparable
to the half-power size of the far-IR continuum source. The very strong continuum flux from the 
central region makes that a significant fraction of the H$_2$O and OH lines toward the Hot Core 
 appear either in absorption or show P-Cygni profiles (Figure~\ref{fig:pacs_profiles}). 
These characteristic line profiles probe the expanding outflows in the region and were
previously inferred from \textit{KAO} and \textit{ISO} observations 
\citep[e.g.,][]{Mel90,Wri00,Cer06,Goi06}.
The much higher angular resolution of \textit{Herschel} allows us to spatially resolve
the regions where these lines switch to pure emission as the far-IR
continuum  becomes weaker, as the gas density increases or both. % (\textit{e.g.,} toward Peak~1).
Figures \ref{fig:orion_maps_pacs}$d$ and \ref{fig:orion_maps_pacs}$e$ show the spatial distribution of the
$o$-H$_2$O 4$_{23}-3_{12}$ ($E_{\rm u}/k$=398~K) and OH \mbox{$^2\Pi_{1/2}$-$^2\Pi_{3/2}$}  \mbox{$J$=1/2-3/2} 
($E_{\rm u}/k$=182~K)  lines at $\sim$79\,$\mu$m.
Both lines show a similar spatial distribution and are representative of other
H$_2$O and OH lines, with the absorption lines peaking toward
the continuum source and the emission lines
showing a NW-SE distribution characteristic of the wide-angle H$_2$ outflow  %high-velocity outflow and
and peaking toward the Peak~1 and 2 regions. 

The [\CII]\,158\,$\mu$m and [\OI]\,145\,$\mu$m lines have also been mapped 
and they show a  different spatial distribution compared to  
molecular lines. Their emission distribution show a more NE-SW orientation with the
[\OI]\,145\,$\mu$m line peaking $\sim$15$''$ NE from the Hot Core.
%The Peak~1 region shows some  [\OI]\,145\,$\mu$m emission that is not seen toward Peak~2.
The [\CII]\,158\,$\mu$m line shows a similar spatial distribution, but the emission toward
Peak~1  and 2 is even weaker.

\begin{figure}[t]
\begin{center}
\includegraphics[angle=0,scale=.37]{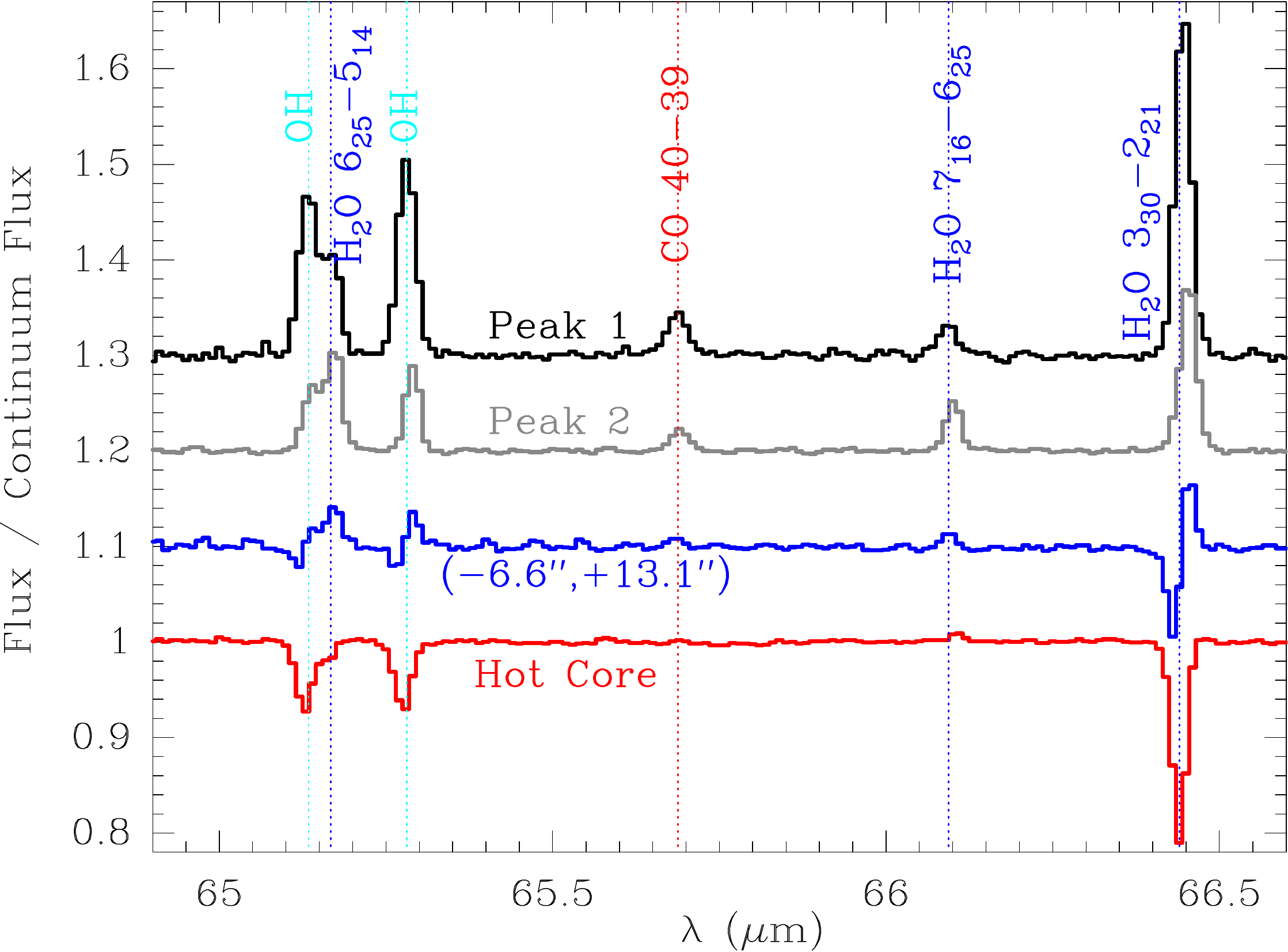} 
\caption{\textit{Herschel}/PACS spectra around $\sim$65\,$\mu$m showing the
evolution of H$_2$O and OH line profiles toward Peak~1 (black), Peak~2 (gray), the Hot Core (red)
and toward an intermediate position (blue). The continuum flux is very strong
toward the Hot Core  and decreases toward Peak~1/2 which show  pure emission far-IR spectra.}
\label{fig:pacs_profiles}
\end{center}
\end{figure}

\subsection{The $\sim$54-310\,$\mu$m spectrum of Orion~Peak~1/2}

Figure \ref{fig:pacs_spire_spectra} shows the pure emission spectrum extracted toward
Peak~1 (black histograms)  and 
Tables~A.1 to A.6 list the fluxes of all detected lines in the 54-310\,$\mu$m range.
The first  complete far-IR spectrum taken toward Orion BN/KL
with \textit{ISO} \citep{Ler06} had a very poor angular resolution of $\sim$80$''$, thus
mixing different physical components and also mixing the emission/absorption components
in the line of sight.
With \textit{Herschel} we can extract and isolate %for the first time 
the complete far-IR spectrum toward H$_{2}$~Peak~1. 
This region is also an emission peak in the very high-$J$ CO
 lines ($J>30$). In the following we describe the main
attributes of the far-IR spectrum toward H$_2$~Peak~1 shown in Figure~\ref{fig:pacs_spire_spectra}. 
 
 More than 100 lines have been detected, most of them rotationally excited lines
from abundant molecules: 34 $^{12}$CO lines (up to $J$=48-47 and $E_{\rm u}$/$k$$=$6458\,K), 
44 lines of both $o$-H$_2$O and  $p$-H$_2$O (up to 8$_{26}$-7$_{35}$ and $E_{\rm u}$/$k$$=$1414\,K), 
13  OH lines (up to $^2\Pi_{1/2}$ $J$=9/2-7/2 and $E_{\rm u}$/$k$$=$876\,K),
11 $^{13}$CO lines (up to $J$=19-18 and $E_{\rm u}$/$k$$=$1004\,K) and several HCN  lines 
(up to $J$=15-14 and $E_{\rm u}$/$k$$=$510\,K).
Several atomic fine structure lines are detected: [\CII]\,158$\mu$m, [\OI]\,63,145$\mu$m,
[\OIII]\,88$\mu$m and [\NIII]\,57\,$\mu$m. Note that the last two lines arise 
from the foreground \HII\,region (the Orion nebula). Hence, a significant contribution of the [\CII] and [\OI] line emission 
arises from the PDR interfaces between this very extended \HII\,region and the molecular cloud that embeds BN/KL
and thus they are not be entirely related with the molecular outflows. 

Approximately 10\% of the observed  H$_2$O and CO emission toward Peak~1 arises from high energy transitions, with 
$E_{\rm u}/k >$600\,K for water vapor and $E_{\rm u}/k >$2000\,K  for CO ($J>$27). 
Here we are specially interested in
constraining the abundances and physical conditions of the material where these high excitation transitions arise.

\begin{figure*}[ht]
\begin{center}
\includegraphics[angle=0,scale=0.8]{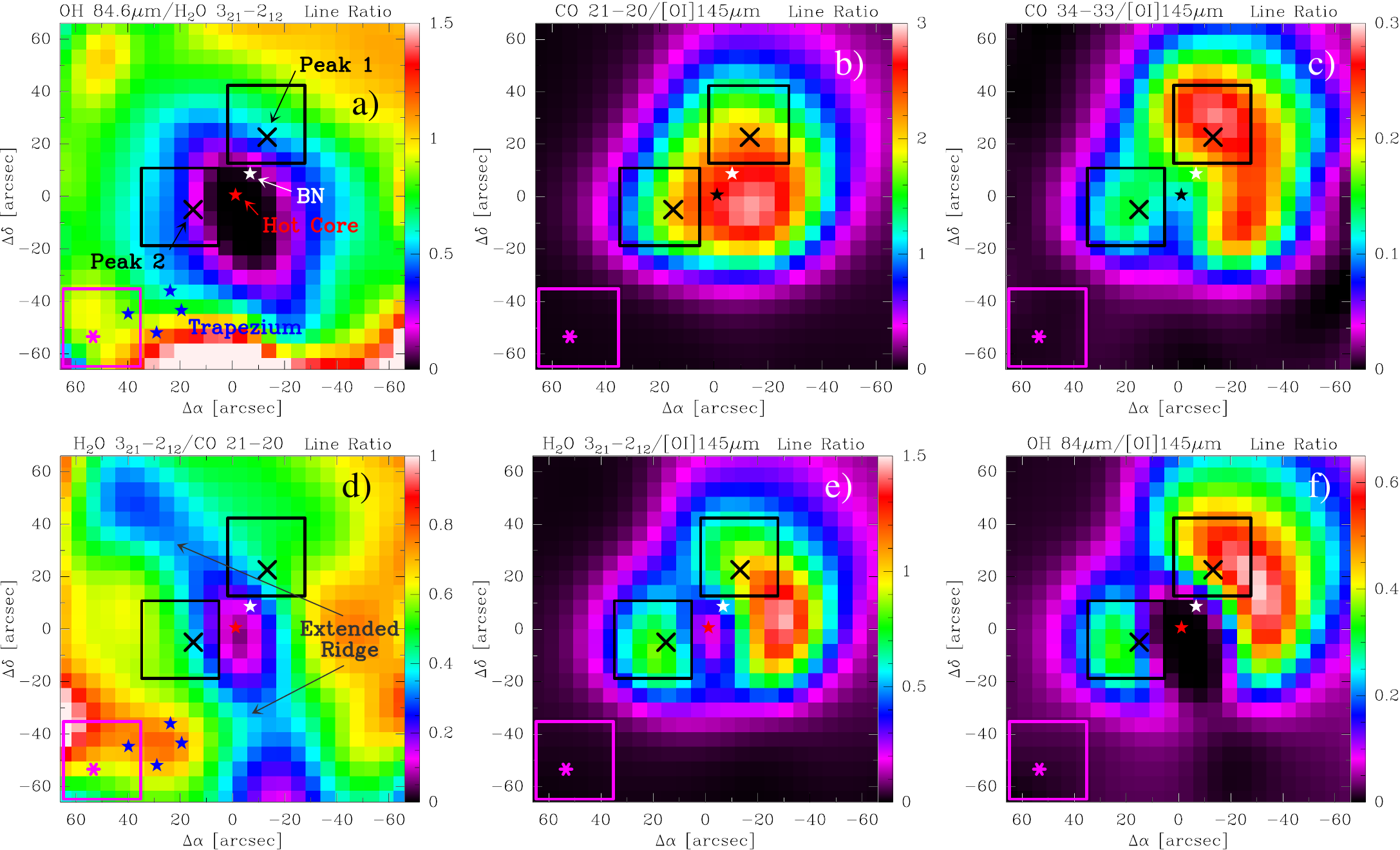} 
\caption{Line surface brightness ratios for selected lines observed with \textit{Herschel}/PACS toward Orion BN/KL.
The areas used to compute the line luminosities in the three representative regions discussed in the text  
(Peak~1, Peak~2 and ambient cloud)  are shown with black and pink squares respectively.
The black crosses and the pink star show the exact positions where the spectra shown in Figure~\ref{fig:pacs_spire_spectra} are extracted. 
The position of the main sources discussed in the text: the Trapezium cluster stars (blue stars), 
the BN source and the position of the Hot Core (colored stars) are also shown.}
\label{fig:observed_ratios}
\end{center}
\end{figure*}

The Peak~2 region displays a similar molecular spectrum with only small differences in the line fluxes.
In particular, Peak~2 shows a factor $\approx$2 weaker CO and OH lines compared to Peak~1. 
Most of the  water vapor lines are also up to a factor $\sim$2 fainter,  although some excited H$_2$O
lines (typically with $E_u$/$k$$>$500\,K) are slightly brighter  towards Peak~2.
A comparison between the Peak~1 and 2 shortest-wavelength  spectra is shown in Figures~\ref{fig:pacs_spire_spectra}$c$,
\ref{fig:pacs_spire_spectra}$d$ and \ref{fig:pacs_spire_spectra}$e$. 
A closer look to the line profile evolution near 65\,$\mu$m is shown in Figure~\ref{fig:pacs_profiles}.

Figure \ref{fig:pacs_spire_spectra} also shows the spectrum toward a position
in the ambient cloud (pink histogram) far from the Hot Core and outflows.
As an example, the $^{12}$CO \mbox{$J$=17-16} line is $\sim$250 times weaker than toward 
Peak~1 and the very \mbox{high-$J$} CO lines are not even detected.
The mid-$J$ $^{13}$CO lines are only marginally detected 
due to the low sensitivity of SPIRE to detect faint and narrow lines. This shows that much lower column densities
of warm CO exist toward the ambient cloud  compared to Orion BN/KL  \citep[see Peng et al.][for 
large-scale $^{13}$CO $J$=8-7 mapping of OMC1 with APEX]{Pen12b}. 
On the other hand, the bright [\OIII]\,88$\mu$m fine structure line from ionized gas shows similar luminosities
in both positions, confirming that this is a different structure in the line-of-sight (the foreground ionized nebula
illuminated by the Trapezium cluster).

\section{Analysis}

\subsection{Maps of line surface brightness ratios}

In addition to the absolute line surface brightness, our multi-line spectral  maps allow one to 
study the spatial variations of different line intensity ratios, all observed with similar angular resolution
and calibration accuracy. 
Figures~\ref{fig:observed_ratios} and \ref{fig:lvg_ratios} show a selection of such maps.

When compared to the [\OI]\,145$\mu$m optically thin line for example, the CO $J$=21-20 and the $J$=34-33 intensity ratio maps
suggest the presence of two different  physical components, a lower excitation component showing a roughly
spherical distribution, and a more excited component that shows
a maximum toward the Peak~1 region (Fig.~\ref{fig:observed_ratios}$c$).
In particular, the emission of the CO lines with  $J$ from $\simeq$20 to 25 peak $\sim$10$''$ NE of the Hot Core
and shows a low  CO $J$=21-20/[\OI]\,145$\mu$m intensity ratio. This may indicate a shocked region where
CO is being dissociated. On the other hand, the CO \mbox{$J$=34-33/[\OI]\,145$\mu$m} intensity ratio map peaks toward
Peak~1, the same region where the very excited CO lines ($J$$>$30) have their intensity peak.
The OH/$o$-H$_2$O (84.6/75.4\,$\mu$m) intensity ratio map (similar Einstein coefficients and upper level energies)
shown in Fig.~\ref{fig:observed_ratios}$a$ suggests that the OH/H$_2$O abundance increases at the outer edge 
of the  outflow(s). Note that, as a consequence of H$_2$O photodissociation, 
the OH/H$_2$O intensity ratio gets even higher toward the southernmost regions
of the map directly illuminated by the strong UV radiation 
from the Trapezium cluster \citep[see][for OH observations toward the Orion Bar PDR]{Goi11}.
Figure~\ref{fig:observed_ratios}$d$ shows the  $o$-H$_2$O~3$_{21}$-2$_{12}$/CO~$J$=21-20 line
ratio map. This ratio traces warm gas regions where the water emission is enhanced.
The map shows the smallest H$_2$O/CO intensity ratios along the NE-SW direction of the quiescent extended ridge and the highest
values in the perpendicular direction (the high-velocity outflow). 

\begin{figure*}[ht]
\begin{center}
\includegraphics[angle=0,scale=.8]{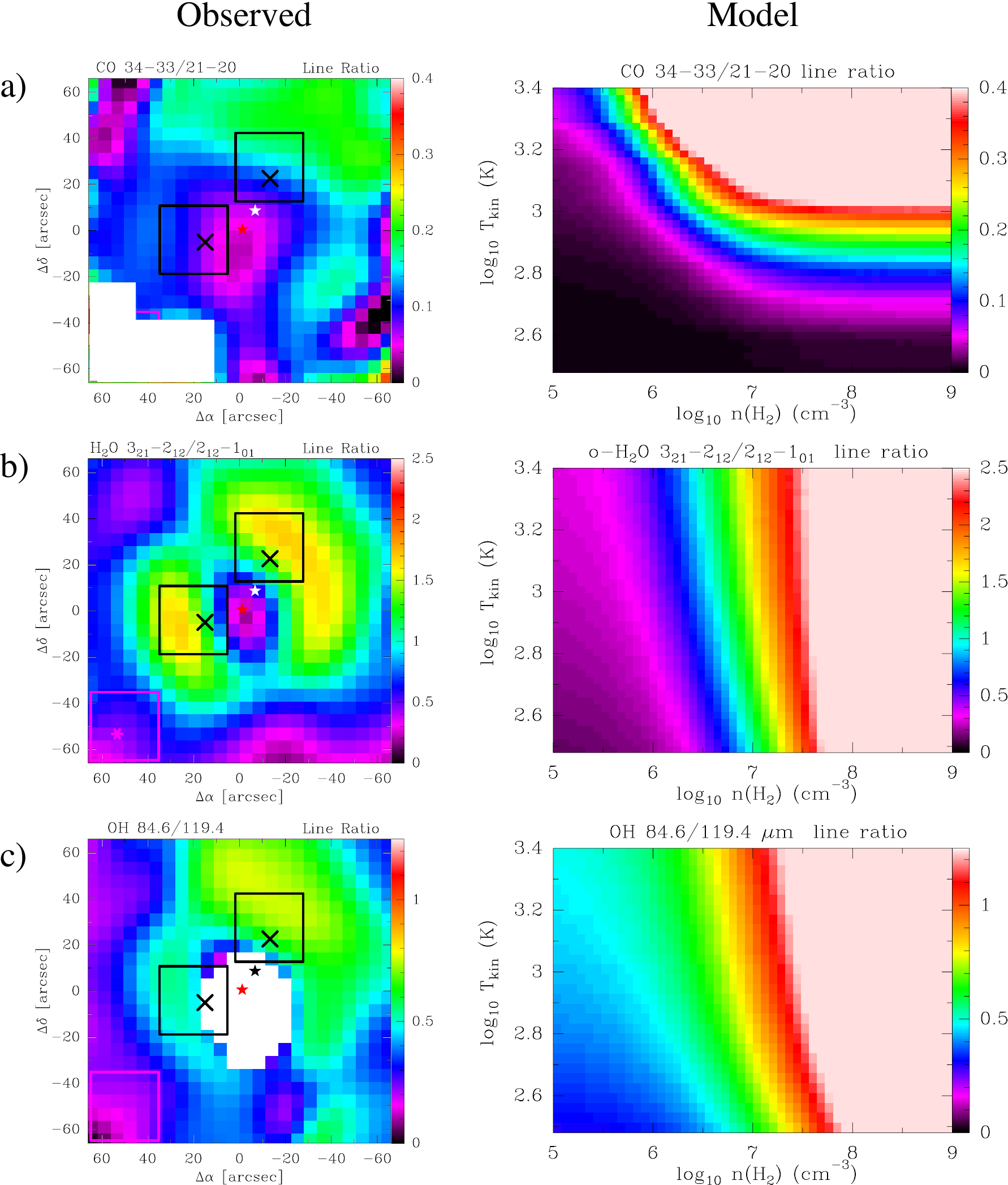} 
\caption{Line surface brightness ratios for selected lines observed with \textit{Herschel}/PACS 
(\textit{left} column) and non-LTE excitation models for the same line intensity ratio (\textit{right} column). 
The areas used to compute the line luminosities in the three representative regions discussed in the text  
(Peak~1, Peak~2 and ambient cloud)  are shown with black and pink squares respectively.
The black crosses and the pink star show the exact positions where the spectra shown in Figure~\ref{fig:pacs_spire_spectra} are extracted. 
The position of  the BN source and the position of the Hot Core  are also shown in each panel (colored stars).
Panels $a)$, $b)$  and $c)$ show the CO $J$=34-33/21-20 (77.1/124.2\,$\mu$m), 
$o$-H$_2$O 3$_{21}$-2$_{12}$/2$_{12}$-1$_{01}$ (75.4/179.5\,$\mu$m)
and OH $^2\Pi_{3/2}$  $J$=7/2-5/2\,/\,5/2-3/2 (84.6/119.4\,$\mu$m) intensity ratios respectively. 
Note that for a given species, the color scale in the  line ratio map and in the grid of models are the same.
The blanked pixels (in white) represent either very low S/N emission (for CO) or line absorption
detections (for OH).} 
\label{fig:lvg_ratios}
\end{center}
\end{figure*}

As a more quantitative (but still approximated) way of investigating the temperature and density gradients
in the region, we selected three appropriate CO, H$_2$O and OH line intensity ratios and used a non-LTE, LVG 
radiative transfer code \citep{Cer12} to explore their variation over a large parameter space.
The assumed column densities follow the observed far-IR line luminosity ratios $L$(CO)/$L$(H$_2$O)/$L$(OH)$\simeq$5/2/1 
and we chose
$N$(OH)=10$^{16}$\,cm$^{-2}$ (see Table~\ref{table:luminosities}).
A line-width of 30\,km\,s$^{-1}$ representative of Orion outflow(s) was used.
The latest available collisional rates were used (Daniel et al. 2011 and references therein for 
H$_2$O,  Yang et al. 2010, extended by Neufeld 2012, for CO and van Offer et al 1994 for OH).
We used an H$_2$ $ortho$-to-$para$ (OTP) ratio of 3, the inferred value from H$_2$ observations
in the region \citep{Ros00}.

Figure~\ref{fig:lvg_ratios} shows the spatial distribution of the 
CO~$J$=34-33/21-20 (77.1/124.2\,$\mu$m),  $o$-H$_2$O \mbox{3$_{21}$-2$_{12}$/2$_{12}$-1$_{01}$} (75.4/179.5\,$\mu$m)
and OH $^2\Pi_{3/2}$  \mbox{$J$=7/2-5/2\,/\,5/2-3/2} (84.6/119.4\,$\mu$m)  intensity ratios 
respectively. The grid of non-LTE  models of the same line ratio is
shown  in the  \textit{right} column. Note that for a given species the color scale in the  intensity ratio
map and in the grid of models are the same.

As a first order, the CO \mbox{$J$=34-33/21-20} intensity ratio traces (warm gas) temperature variations 
in the region. Figure~\ref{fig:lvg_ratios}$a$ shows that the intensity ratio increases with distance to
the Hot Core, reaching a maximum value in the NW direction near Peak~1 where the high-velocity outflow
plunges into the ambient cloud. These shocked gas regions are characterised by high 
temperatures (at least $T_{\rm k}$$>$800\,K).  
Note that a much lower  CO \mbox{34-33/21-20} ratio is inferred toward the inner Hot Core regions.
However, owing to the presumably high far-IR dust opacities toward the Hot Core (see Sect.~4.3), it is difficult to conclude whether
the elevated dust opacity attenuates the high-$J$ CO line emission at the shortest far-IR wavelengths or this is just the
absence of hot shocked gas toward the Hot Core. The lack of high-velocity CO bullets in the Hot Core region, however, 
seems to support this latter scenario \citep[see the interferometric
observations by][]{Zap09,Zap11}.

\vspace{0.5cm}

The selected H$_2$O and OH line ratios are more sensitive to density variations in  the shocked gas.
Both Figs.~\ref{fig:lvg_ratios}$a$ and \ref{fig:lvg_ratios}$b$ suggest that the gas density increases toward Peak~1 and 2.
In these positions the gas seems compressed by a factor $\gtrsim$10, reaching
$n$(H$_2$)$\approx$10$^7$\,cm$^{-3}$. Still, note that the  $T_{\rm k}$ and $n$(H$_2$) values
suggested in Figure~\ref{fig:lvg_ratios} are merely indicative and only show global trends in the region. 
The exact values will depend on the actual column density of material toward each line-of-sight, the local
excitation conditions and radiative transfer effects. 

Finally, the line ratio maps shown in
Figures~\ref{fig:observed_ratios}$a$, \ref{fig:observed_ratios}$d$ and \ref{fig:lvg_ratios} 
suggest that there are not major differences between the physical conditions prevailing in the Peak~1 region
and those of Peak~2 (see  also sec. 4.2).
In the following, we focus our analysis in the interpretation of the spectra toward Peak~1.

\begin{figure}[ht]
\begin{center}
\includegraphics[angle=0,scale=1.0]{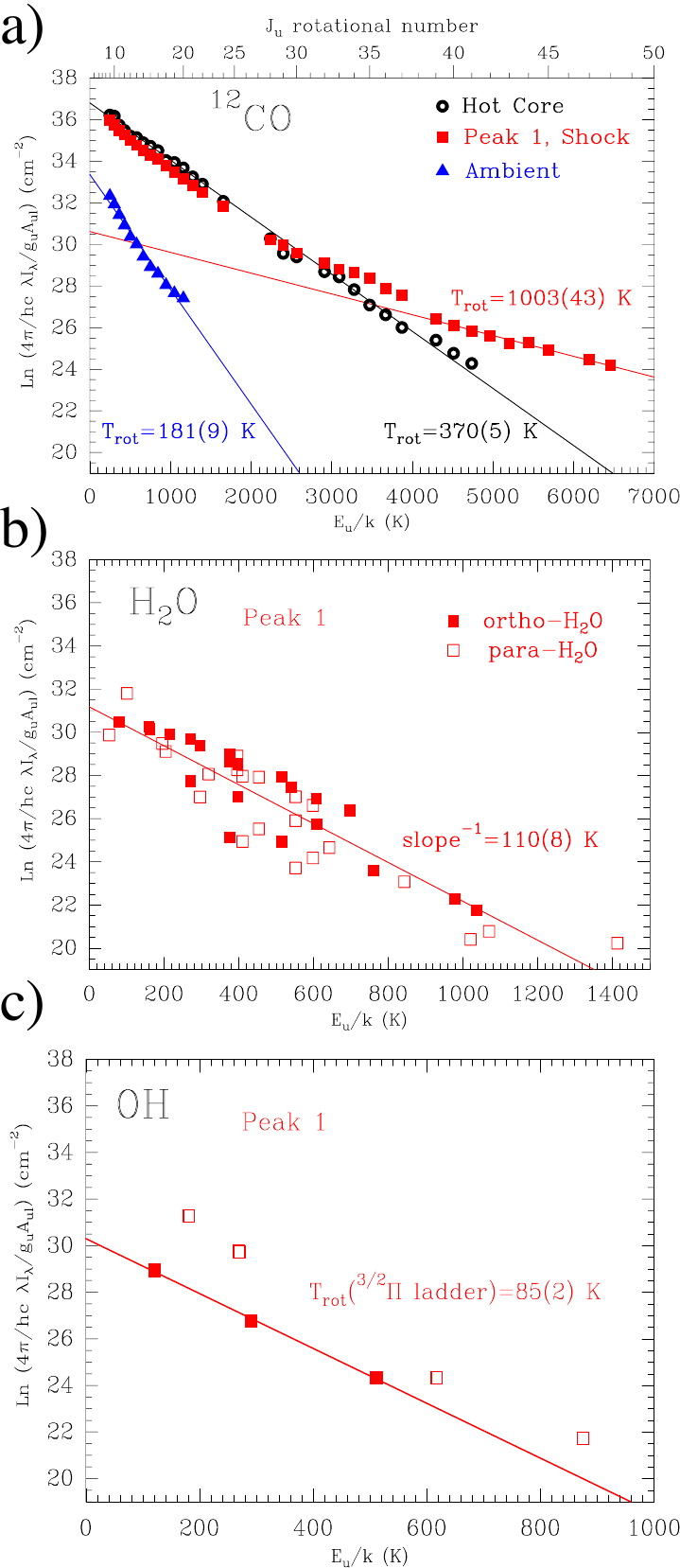} 
\caption{Rotational population diagrams obtained from \textit{Herschel}/PACS and SPIRE observations
of the Peak~1 region in a $\sim$30$''$$\times$30$''$ aperture.
$a)$ $^{12}$CO rotational diagram also showing the corresponding data toward the 
ambient cloud (shown in Figure~1 with a pink square) and toward the Hot Core, all in apertures of the same size. 
The straight lines and
associated rotational temperatures show fits to the data set in different $J$ ranges.
$b)$ $o$- and $p$-H$_2$O rotational diagram. A single straight line is fitted to all points.
$c)$ OH rotational diagram. $T_{\rm rot}$ is calculated by fitting OH lines
in the $^2\Pi_{3/2}$ rotational ladder only.}
\label{fig:rot_diagrams}
\end{center}
\end{figure}

\subsection{CO, H$_2$O and OH rotational diagrams}

Figure~\ref{fig:rot_diagrams} shows all CO,  
H$_2$O and OH detected lines toward Peak~1 in a \textit{rotational diagram}.
In these plots we assumed extended emission over a $\sim$30$''$$\times$30$''$ aperture.
Owing to the high densities previously inferred toward Peak~1,
$\gtrsim$$10^{6}$\,cm$^{-3}$ from H$_2$ line observations \citep{Ros00} and $\gtrsim$$10^{7}$\,cm$^{-3}$ from CO
and H$_2$O ro-vibrational lines \citep{Gon02}, the rotational temperatures ($T_{\rm rot}$) derived from the CO 
rotational diagrams are a good lower limit to the gas temperature ($T_{\rm k}$$\gtrsim$$T_{\rm rot}$).

The  $^{12}$CO rotational diagram 
shows a notable contribution of very  high-$J$ lines ($J$$>$40) that cannot be fitted with
a single rotational temperature in LTE (a straight line in the plot). 
A smilar population diagram of the  mid-IR H$_2$ lines also shows
a remarkable positive curvature \citep{Ros00}. Mid-IR H$_2$ rotational lines in the ground-vibrational 
state (up to $E_{\rm u}$/$k$$\lesssim$10$^4$\,K) have low critical densities, of the order of
 10$^{2-6}$\,cm$^{-3}$, and thus the  curved
H$_2$ level  distribution is not compatible with a low-density isothermal
component (see also Sect.~6.2). 
As H$_2$, the  $^{12}$CO rotational diagram suggests the presence of at least two temperatures components,
a warm component with $T_{\rm rot}$$\approx$400\,K and a hotter component with 
$T_{\rm rot}$$\approx$1000\,K. %that dominates the CO line emission with $J$$>$40. 
For comparison, the CO emission in the ambient cloud is dominated by a cooler component
with $T_{\rm rot}$$\approx$180\,K (blue triangles in Fig.~\ref{fig:rot_diagrams}$a$). 
This is the warm and extended  face-on PDR (the interface between the foreground \HII\,region and the molecular cloud),
heated by  FUV photons from the Trapezium stars \citep[see Peng et al.][for large-scale mapping of several mid-$J$ CO narrow lines]{Pen12b}.

A  rotational diagram of the $^{13}$CO lines detected with PACS toward Peak~1 provides 
$T_{\rm rot}$$\approx$200\,K, thus  lower than the $T_{\rm rot}$ inferred from $^{12}$CO. 
The measured  $^{12}$CO/$^{13}$CO intensity ratio for the $J$=16-15 and 15-14 lines is
55$\pm$5, lower than the  $^{12}$C/$^{13}$C isoptopic ratio of $\sim$79$\pm$7 in Orion\citep{Lan90}.
For $J$$>$16, however,  the $^{12}$CO/$^{13}$CO intensity ratio becomes equal or larger than the isotopic value,
showing that while the submm $^{12}$CO low- and mid-$J$  lines are optically thick, 
the high-$J$ far-IR  lines ($J$$>$16) are unambiguously optically thin.

H$_2$O and OH have a more complex rotational spectroscopy and much higher critical densities than 
H$_2$ and CO lines. 
Therefore, the rotational temperatures inferred from their rotational population diagrams
do not represent $T_{\rm k}$ properly. We show them in Fig.~\ref{fig:rot_diagrams}$b$ and 
\ref{fig:rot_diagrams}$c$ so that
they can be compared with similar diagrams published in the literature 
\citep[e.g.,][and references therein]{Kar14}. 
H$_2$O and OH rotational levels are subthermally populated 
($T_{\rm rot}$$<$$T_{\rm k}$) and in regions of strong far-IR continuum fields
like the Orion Hot Core,
they are affected by radiative pumping \citep[$T_{\rm rot}$$<$$T_{\rm d}$; 
see e.g.,][]{Goi06,Cer06}.

\subsection{Simple non-LTE model of Peak~1}

In order to constrain the average physical conditions and  abundances in
the different temperature components suggested by the CO, H$_2$O and OH rotational diagrams,
we have carried out a simple non-LTE model of the observed emission toward Peak~1.

Our model  includes radiative pumping by far-IR dust continuum photons. In particular,
we add the effects of a modified black body source with a radius of 10$''$ and a dust color 
temperature of 100\,K. The continuum source is optically thick at 100\,$\mu$m 
(with $\tau_{\lambda}$=7\,$[$100/$\lambda$($\mu$m)$]$$^2$). These values were obtained by
fitting the far-IR PACS continuum and the  mid-IR photometric observations 
obtained by \textit{SOFIA}/FORECAST  in a $\sim$30$''$$\times$30$''$ aperture
around the Hot Core and IRc sources \citep{Bui12}. The modelled gas components were placed at 25$''$ from the continuum source.

We follow previous studies of the  CO and H$_2$O rotational and ro-vibrational  emission observed with $\textit{ISO}$. % \citep{Gon02}.
In particular, a detailed excitation analysis of the  CO \mbox{$v$=1-0} and 
 H$_2$O $v_2$=1-0  bands
\citep[observed with the 14$''$$\times$20$''$ aperture of $\textit{ISO}$/SWS;][]{Gon02} 
concluded that the CO vibrational band at $\sim$4.7\,$\mu$m  
can be reproduced with two dense  ($n$(H$_2$)$\sim$2$\times$10$^{7}$\,cm$^{-3}$) temperature components of shock-excited material; 
a \textit{warm} one with $T_{\rm k}$$\sim$200-400\,K and a \textit{hot} component with $T_{\rm k}$$\sim$3000\,K.
This high density is in principle consistent with the compression factors suggested by our intensity ratio maps
(Fig.~\ref{fig:lvg_ratios}$b$ and \ref{fig:lvg_ratios}$c$) and with the presence of excited emission lines from high density tracers like HCN.

In our model, the \textit{hot} and \textit{warm} components have a size of 10$''$,
consistent with the compact spatial distribution of the very high-$J$ CO lines  toward Peak~1
\citep[see Fig.~\ref{fig:orion_maps_pacs}$g$ and][]{Sem00}.
Figure~\ref{fig:co_lvg_mods} shows the observed CO line fluxes   
as a function of $J$. The distribution of line fluxes shows 
a maximum at $J$=17 and an emission tail at $J$$>$40.
The CO lines with $J$$<$20 show a more extended spatial distribution 
than the higher-$J$ lines (Fig.~\ref{fig:orion_maps_pacs}$i$) suggesting
that a third, \textit{cool} and extended 
component needs to be included to fit the CO lines below $J$=20.
We adopt a line-width of 30~km\,s$^{-1}$ in all modelled components.
This is consistent with the wide-angle high-velocity outflow and with the typical
line-widths of CO, H$_2$O and many other species 
observed at heterodyne spectral resolution in the plateau  \citep[e.g.,][]{Bla87,Mel10,Ter10}.

\begin{table*}[t]
      \caption[]{Model components  and resulting source-averaged column densities toward Peak~1
(assuming a line-width of 30\,km\,s$^{-1}$).\\\ $\dagger$Assuming no contribution in the \textit{hot} component (see text).}
\centering
\begin{tabular}{lcccccc}
\hline\hline
\multicolumn{1}{c}{} & 
\multicolumn{1}{c}{$T_{\rm k}$ } &
\multicolumn{1}{c}{$n$(H$_2$)} &
\multicolumn{1}{c}{Size} &
\multicolumn{1}{c}{$N$(CO)} &
\multicolumn{1}{c}{$N$(H$_2$O)} &
\multicolumn{1}{c}{$N$(OH)} \\
 Component & (K) & (cm$^{-3}$) & (arcsec) & (cm$^{-2}$) & (cm$^{-2}$) & (cm$^{-2}$) \\\hline\hline
\textit{Hot} shock-excited       & $\sim$2500 & $\sim$2$\times$10$^7$ & $\sim$10 & $\sim$1.5$\times$10$^{16}$ & $\sim$2.0$\times$10$^{16}$  & $\sim$1.5$\times$10$^{16}$\\
\textit{Warm} shock-excited      & $\sim$500  & $\sim$2$\times$10$^7$ & $\sim$10 & $\sim$2.0$\times$10$^{19}$ & ($\sim$2.0$\times$10$^{17}$)$^\dagger$ & ($\sim$5.0$\times$10$^{16}$)$^\dagger$ \\
\textit{Cool} plateau outflow(s) & $\sim$200  & $\sim$2$\times$10$^6$ & $\sim$30 & $\sim$2.0$\times$10$^{19}$ & $\sim$6.7$\times$10$^{17}$ & $\sim$1.5$\times$10$^{17}$\\\hline\hline
\end{tabular}
\vspace{0.2cm}
\label{table:model_components}
\end{table*}

\vspace{0.2cm}
\begin{figure}[b]
\begin{center}
\includegraphics[angle=0,scale=.5]{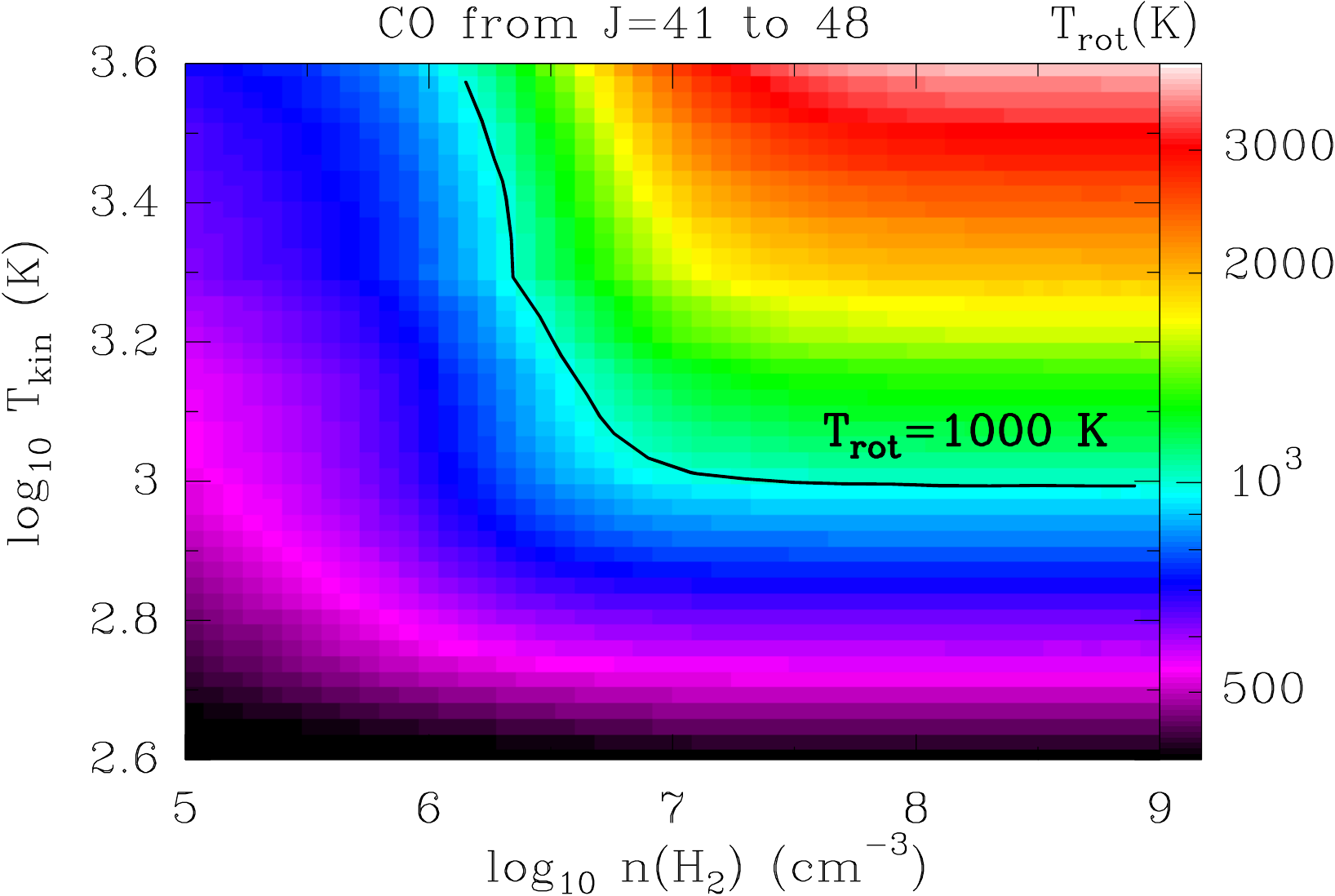} 
\caption{Synthetic $^{12}$CO rotational temperatures ($J$= 41-48 range) obtained from a grid of 
single-temperature non-LTE models with $N$(CO)=5$\times$10$^{16}$\,cm$^{-2}$. % and $\Delta$v=30\,km\,s$^{-1}$. 
The $T_{\rm rot}$=1000\,K isocontour is shown as a black curve.
This is the  $T_{\rm rot}$ inferred from the very high-$J$ CO lines observed with PACS 
 toward Peak~1.}
\label{fig:hot_co_trot}
\end{center}
\end{figure}

\begin{figure}[b]
\begin{center}
\includegraphics[angle=0,scale=.47]{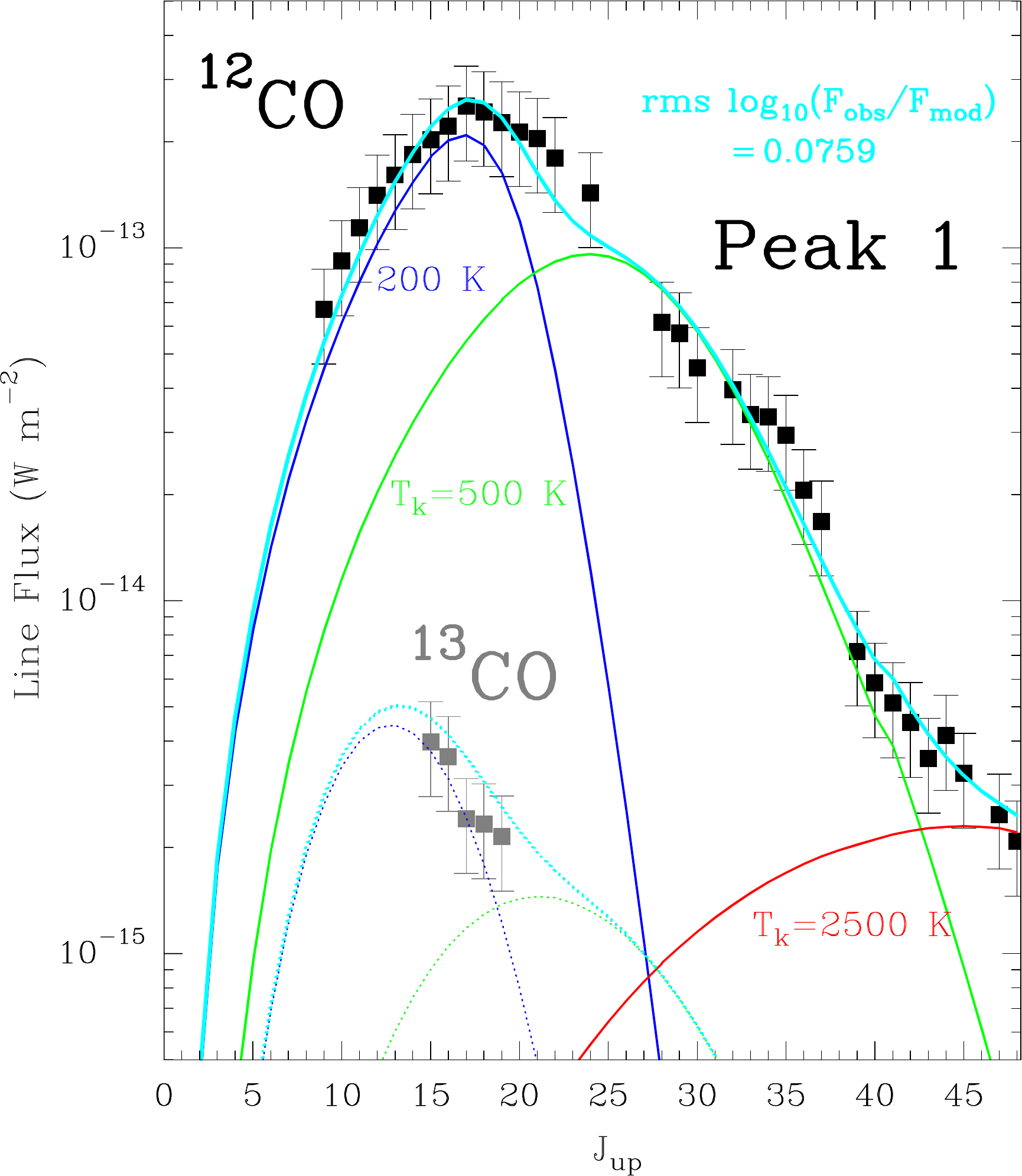} 
\caption{$^{12}$CO and $^{13}$CO line fluxes observed by \textit{Herschel}/PACS and SPIRE
toward Orion~Peak~1 and non-LTE excitation and radiative transfer model results
(continuos and dashed curves for $^{12}$CO and $^{13}$CO respectively). 
The blue curves represent the contribution of the $\sim$200\,K (\textit{cool} plateau) component,
the green curves represent the $\sim$500\,K (\textit{warm} shock-excited) component and
the red curve represents the   $\sim$2500\,K (\textit{hot} shock-excited) component.
 The cyan curves show the result of adding all modeled temperature components.}
\label{fig:co_lvg_mods}
\end{center}
\end{figure}

\subsubsection{$^{12}$CO and $^{13}$CO model}

The presence of \textit{hot} gas  with $T_{\rm k}$$>$1000\,K was 
first inferred from  H$_2$  ro-vibrational temperatures \citep[][]{Bec78,Ros00}
and is also needed to fit the high energy CO $v$=1-0 ro-vibrational emission \citep{Gon02}.  
In the far-IR spectrum, this component becomes apparent in the CO $J$$>$40 ($E_{\rm u}/k>4513$\,K)
emission tail. 
Figure~\ref{fig:hot_co_trot} shows synthetic $^{12}$CO rotational temperatures in the $J$= 41-48 range obtained 
from a grid of single-temperature non-LTE models.
It shows that for densities in the range $n$(H$_2$)$\simeq$10$^{6.2}$ to 10$^{8}$\,cm$^{-3}$,
the gas temperatures required to produce $T_{\rm rot}$(CO)$\simeq$1000\,K for $J$$>$40 are very high, in the range 
$T_{\rm k}$$\simeq$3500 to 1000\,K respectively.

The \textit{warm} and \textit{hot} components dominate
the CO line emission for $J>20$. Assuming the same H$_2$ densities 
inferred from the CO $v$=1-0 band analysis, we run models varying $T_{\rm k}$ and $N$(CO)
around the values derived by Gonz{\'a}lez-Alfonso et al. (2002) and Sempere et al. (2000). 
A good fit to the data is found for  $T_{\rm k,\,warm}$$\simeq$500\,K and $T_{\rm k,\,hot}$$\simeq$2500\,K. 
The inferred source-averaged CO column densities are  similar to those inferred 
from the CO $v$=1-0 vibrational band:
$N$(CO)$_{\rm warm}\simeq2.0\times10^{19}$\,cm$^{-2}$ and $N$(CO)$_{\rm hot}\simeq1.5\times10^{16}$\,cm$^{-2}$.

A third \textit{cooler} and more extended component is needed to reproduce the $^{12}$CO and $^{13}$CO 
emission with $J$$<$20. We associate this component with the extended plateau  (outflowing  and swept-up material). 
The gas  density in the plateau is estimated to be $n$(H$_2$)$\approx$(1-2)$\times$10$^6$\,cm$^{-3}$
\citep[e.g.,][]{Bla87,Mel10,Ter10}. This is, at least a factor 10 lower than in the shock-excited gas
 traced  by the very high-$J$ CO rotational lines  
and by the CO $v$=1-0  band (\textit{hot} and \textit{warm} components). 
Modelling all components together (adding the emission from each component)
we obtain a satisfactory  fit to the CO ladder with $T_{\rm k,\,plat}$$\simeq$200\,K, 
$n$(H$_2$)$\simeq$2$\times$10$^6$\,cm$^{-3}$ and $N$(CO)$_{\rm plat}$$\simeq$2$\times$10$^{19}$\,cm$^{-2}$
for the \textit{cool} component (Fig.~\ref{fig:co_lvg_mods}). 
The complete model agrees with observations within a $\sim$20\,\% and 
is also consistent with the  $^{13}$CO lines assuming a $^{12}$C/$^{13}$C isotopic ratio of 80.

\begin{figure*}[]
\begin{center}
\includegraphics[angle=0,scale=.76]{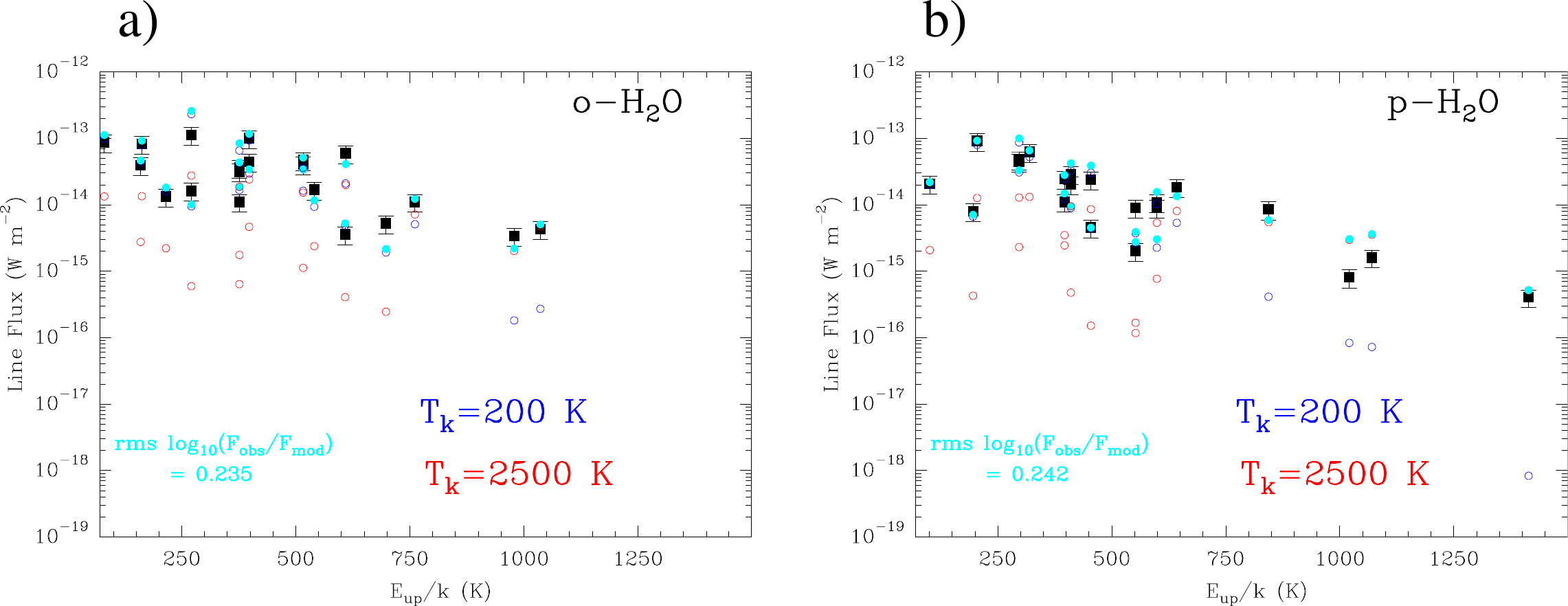} 
\caption{$o$-H$_2$O and $p$-H$_2$O line fluxes observed by \textit{Herschel}/PACS and SPIRE
 toward Orion~Peak~1 and model results.
 The cyan dots are a model adding the two temperature components (see text).}
\label{fig:h2o_lvg_mods}
\end{center}
\end{figure*}

\subsubsection{H$_2$O model}

The extended plateau  in Orion~BN/KL shows enhanced H$_2$O column densities compared to more quiescent 
interstellar environments.
Still, the exact water vapor abundances are controversial
(\textit{cf.}~Cernicharo et al. 1994, 1999, 2006; Harwit et al. 1998; Wright et al. 2000; Melnick et al. 2000, 2010).
Most of these observations were carried out toward the core of Orion BN/KL where a very strong
mid- and far-IR radiation field from the Hot Core region exists.
The angular resolution of the above observations range from the $\sim$3$'$ and $\sim$80$''$ of \textit{SWAS}
and the \textit{ISO}/LWS, to the 44$''$-12$''$ beam of \textit{Herschel}/HIFI in the submm.
The $\sim$12$''$ resolution of $\textit{Herschel}$/PACS allows us to resolve the intrinsic H$_2$O emission 
 toward Peak~1. In addition, the continuum emission is significantly weaker and 
lines are not contaminated by absorption toward the Hot Core.

%In our model of the far-IR H$_2$O emission we assume the same physical conditions inferred from CO in the previous section.
A single temperature component can not fit the observed H$_2$O emission lines toward Peak~1.
However, most of the H$_2$O lines with $E_{\rm k}/k$$<$750\,K can be reasonably fitted with the
extended plateau component ($T_{\rm k,\,plat}$$\simeq$200\,K, $n$(H$_2$)$\simeq$2$\times$10$^6$\,cm$^{-3}$
and $N$(H$_2$O)$_{\rm plat}$$\simeq$6.7$\times$10$^{17}$\,cm$^{-2}$).
This agrees with detailed models of H$_2$O lines detected by \textit{Herschel}/HIFI toward
the Hot Core position \citep{Mel10}. 

Even taking into account the moderate far-IR radiation field toward Peak~1, the H$_2$O lines arising 
from higher energy levels require higher excitation conditions.
The detailed excitation analysis of the H$_2$O $v$=1-0 vibrational band at $\sim$6.7\,$\mu$m  toward Peak~1 concluded
that hotter and denser gas contributes to the mid-IR H$_2$O ro-vibrational emission.
Since abundant water vapor is expected in the hot excited-gas, we included
H$_2$O in the  \textit{hot} CO component  to match the water emission
lines with $E_{\rm k}/k$$>$750\,K.
A good fit is obtained for 
$N$(H$_2$O)$_{\rm hot}$$\simeq$2$\times$10$^{16}$\,cm$^{-2}$.
Figure~\ref{fig:h2o_lvg_mods} shows that the lower excitation extended plateau component (blue dots)
does not contribute much to the higher energy H$_2$O line fluxes.
On the other hand, the \textit{hot} water component (red dots) contributes to a small fraction of the 
low excitation H$_2$O line fluxes.  In the \textit{hot} component we infer a high H$_2$O/CO$\simeq$1.3 abundance ratio.
Given the simplicity of our model (no radiative coupling between the different components),
the global factor $<$2 agreement is reasonable. The best fit is compatible with an H$_2$O OTP ratio of 3.
For completeness we also computed the maximum allowable H$_2$O column density in the \textit{warm} component ($T_{\rm k}$$\simeq$500\,K)
if the high excitation H$_2$O lines would not arise from the \textit{hot} component ($T_{\rm k}$$\simeq$2500\,K). 
We find $N_{\rm warm}$(H$_2$O)$\simeq$2$\times$10$^{17}$\,cm$^{-2}$, which results in a (modest) maximum H$_2$O/CO$\leq$10$^{-2}$ abundance ratio.
This suggests that high postshock temperatures $>$500\,K are needed to produce high water vapor abundances.

\subsubsection{OH model}
 
Previous OH low-angular resolution far-IR observations 
with $\textit{KAO}$ ($\sim$33$''$) and $\textit{ISO}$ ($\sim$80$''$) unambiguously showed the presence OH in the outflows
expanding from Orion BN/KL (the extended plateau component). This is inferred from the 
observed line-widths and the P-Cygni profiles \citep{Bet89,Mel90,Goi06}.
Our higher angular resolution observations  show a pure absorption OH line spectrum 
toward the Hot Core  (Fig.~\ref{fig:pacs_profiles}) except for the $\sim$163\,$\mu$m doublet that is mainly  
excited by far-IR  continuum photons. The absorbing regions, with a size of 
$\sim$15$''$$\times$20$''$ around the strong continuum peak (see maps in Fig.~\ref{fig:orion_maps_pacs})
reduce the OH line emission fluxes when observed at low angular and spectral resolution. 
Outside the strong far-IR continuum emission regions, all OH lines switch to emission  and they peak
toward the bright H$_{2}$ Peaks~1 and 2.

As H$_2$O, a single temperature component can not reproduce the observed OH emission toward Peak~1.
Following Goicoechea et al. (2006), we model the lower energy OH transitions in the extended plateau.
A satisfactory fit is obtained for 
$N$(OH)$_{\rm plat}$$\simeq$(1.0-1.5)$\times$10$^{17}$\,cm$^{-2}$ 
(or a $N$(OH)/$N$(H$_2$O)$\simeq$0.15-0.22).
However, this model underestimate the fluxes of the OH higher energy transitions.
Such lines likely arise from more excited gas,  
simulated here by the   \textit{warm} and \textit{hot}  components.

\begin{figure}[h]
\begin{center}
\includegraphics[angle=0,scale=.4]{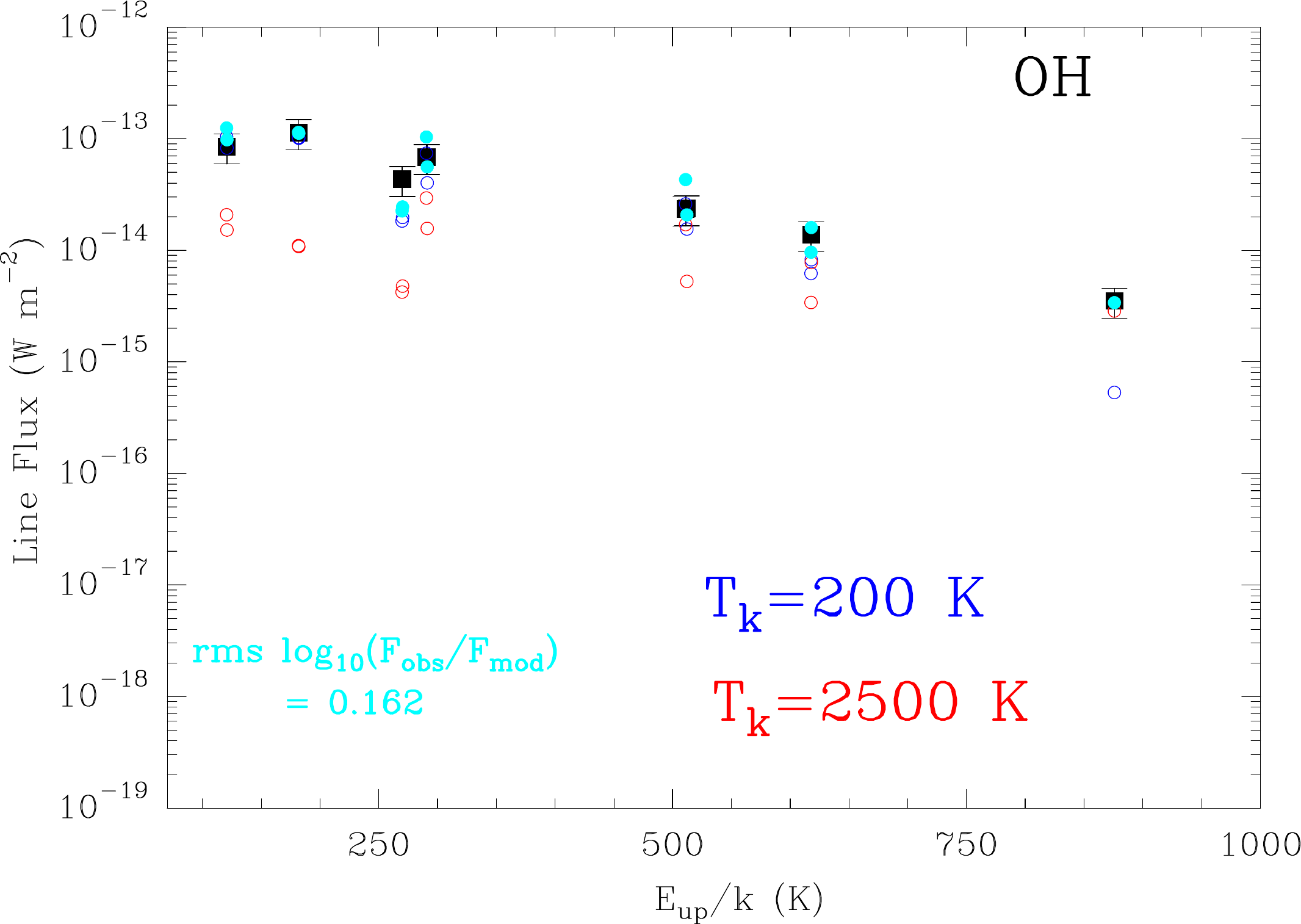} 
\caption{OH line fluxes observed by \textit{Herschel}/PACS and SPIRE
 toward Orion~Peak~1 and model results.
 The cyan dots are a model adding the two temperature components (see text).}
%\vspace{-0.5cm}
\label{fig:oh_lvg_mods}
\end{center}
\end{figure}

\vspace{0.2cm}
We find that if the excited OH emission mainly arises from the \textit{warm}
CO component at $T_{\rm k,\,warm}$$\simeq$500\,K, then
a OH/CO$\simeq$4$\times$10$^{-3}$ column ratio satisfactorily reproduces the  excited OH lines.
On the other hand, if the excited OH lines  arise from the \textit{hot} CO  component
at $\simeq$2500\,K, then a model with roughly equal amounts of hot CO, H$_2$O and OH are needed to fit
the  excited lines.
All in all, the combined extended plateau~+~\textit{hot} or +~\textit{warm} OH model reasonably agrees with observations
within a $\sim$20\%~(Fig.~\ref{fig:oh_lvg_mods}).
Taking into account both scenarios, the source-averaged OH column density in the dense shock-excited component is
$N$(OH)$\simeq$(1.5-5.0)$\times$10$^{16}$\,cm$^{-2}$. % (the lower value for the hotter component).

\subsection{Molecular Abundances toward Peak~1}

The absolute water vapor abundance with respect to H$_2$ has long been discussed in many different protostellar environments.
The old prediction that copious amounts of H$_2$O, and to a lower extent O$_2$, will be produced behind molecular shocks
has been challenged by several \textit{Herschel} observations \citep[e.g.,][]{Gol11,vD13}.
Unfortunatly, even in the $\sim$10-40$''$ beam of \textit{Herschel}, different physical components usually overlap and it is not obvious
to extract their relative abundances with accuracy \citep[e.g.,][]{Che14}. 

Taking $N$(H$_2$)$\approx$10$^{23}$\,cm$^{-2}$ in a 30$''$ beam for the \textit{cool} plateau component
  \citep[][]{Bla87,Mel10}, we obtain $\chi_{\rm plat}$(CO)$\approx$2$\times$10$^{-4}$, $\chi_{\rm plat}$(H$_2$O)$\approx$0.7$\times$10$^{-5}$
and $\chi_{\rm plat}$(OH)$\approx$0.2$\times$10$^{-5}$ abundances. Hence, in the less excited swept-up gas,
water vapor is not the dominant O-bearing species. However, if we only consider
the \textit{hot} shock-excited component, 
the derived  H$_2$O/CO abundance ratio is much higher ($\sim$1.3). Rosenthal et al. (2000) inferred that the column density of hot H$_2$
with T$_{ex}$$>$1200\,K toward Peak~1  is $N$(H$_2$)$\lesssim$10$^{20}$\,cm$^{-2}$. 
Taking the columns from this component alone, %(Table \ref{table:model_components}), 
we  determine  $\chi_{\rm hot}$(H$_2$O)$\gtrsim$2$\times$10$^{-4}$,$\chi_{\rm hot}$(CO)$\gtrsim$1.5$\times$10$^{-4}$ and 
$\chi_{\rm hot}$(OH)$\gtrsim$0.8$\times$10$^{-5}$. 

\subsection{Model uncertainties and Peak~1/2 differences}

Our  multi-component model satisfactorily reproduces the fluxes of most observed lines.
Despite  this model solution is certently not unique, the agreement with observations suggests that the model captures 
the average physical conditions of the shocked-exited gas.

The best fit to the CO $J$=41 to 48 absolute line fluxes  alone  gives $T_{\rm k}$$\simeq$2300-3000\,K and $n$(H$_2$)$\simeq$(3-4)$\times$10$^6$\,cm$^{-3}$
for $N$(CO)=(5.0-2.5)$\times$10$^{16}$\,cm$^{-2}$ for Orion~Peak~1. 
These values suggest that the postshock gas where the high-excitation CO, H$_2$O and OH lines arise might be up to $\sim$5 times
less dense than our assumed value of $n$(H$_2$)$\simeq$2$\times$10$^7$\,cm$^{-3}$
\citep[but see][]{Sem00,Gon02}. In such a case,  the H$_2$O column densities in the \textit{hot} component will be a factor $\sim$4 higher
and the H$_2$O/CO abundance ratio will increase from $\sim$1.3 to $\sim$1.6.
Finally, the presence of a moderate far-IR dust radiation field helps to better fit
the fluxes of several H$_2$O and OH lines. 
Although the radiative pumping effects toward Peak~1  are less
dramatic than toward the Hot Core, neglecting the presence of a far-IR field will overestimate the
the resulting $N$(OH) and $N$(H$_2$O) by  $\sim$50\%.
These numbers should be considered as the range of uncertainty in the physical conditions 
from our simple model (up to a factor $\sim$5 in the absolute volume and column densities).

Our data show small differences between Peak~1 and 2. The CO and OH rotational temperatures we infer toward  Peak~2 
are similar to those toward Peak~1 (despite  lines are up to a factor $\sim$2 brighter).
Only some excited H$_2$O lines are slightly stronger toward Peak~2 (producing higher H$_2$O rotational 
temperatures by $\lesssim$20\%) and
suggesting that the water vapor column and probably the gas density are also slightly higher. 
Note that the same conclusions were reached from the observation of
the CO and H$_2$O vibrational bands \citep{Gon02}.
However, we estimate that the aperture-averaged column densities  and the physical conditions 
differences between both peaks are within a factor $\sim$2 (see also  Figure~\ref{fig:oh_lvg_mods}). 
Therefore, they are smaller than the range of uncertainties in our models
and thus we have not modeled both peaks independently. 

%\vspace{0.5cm}

\section{Gas cooling in Orion H$_2$ Peak~1 shock}

Table~\ref{table:luminosities} summarizes the observed line and continuum luminosities
in a $\sim$30$''$$\times$30$''$ aperture toward the H$_2$~Peak~1 position.
They provide a good measurement of the cooling budget in shocked gas. 
This tabulation also includes an extrapolation of the mid-IR H$_2$ line fluxes
and of the CO and H$_2$O ro-vibrational emission detected by \textit{ISO}/SWS \citep{Ros00,Gon02}.
Their measured luminosities have been multiplied by \\(30$''$$\times$30$''$)/(14$''$$\times$20$''$)$\simeq$3
to approximately take into account the different aperture sizes.
Roughly half of the total line luminosity is provided by 
H$_2$ rotational and vibrational line cooling.
CO is the next most important species, providing $\sim$(22+7)\%\, of the total line luminosity (far\,+\,mid-IR).
H$_2$O only contributes with $\sim$(8+2)\%\,and this may sound surprising in light of previous shock model 
expectations for Orion \citep[e.g.,][]{Dra82,Kau96}.
[\OI] and OH lines contribute to line cooling in similar amounts:
with $L$(\OI)/$L$(H$_2$O)$\simeq$0.4 and  $L$(OH)/$L$(H$_2$O)$\simeq$0.5. 
Taking into account the far-IR range alone, CO rotational lines clearly dominate the gas cooling toward Peak~1 with a very high $L_{\rm FIR}$(CO)/FIR luminosity
ratio of $\approx$5$\times$10$^{-3}$. Note that the  typical luminosity ratio in bright PDRs is below a few 10$^{-4}$ 
(the value in the Orion Bar, Joblin et al. 2015).
In addition, the estimated dust temperatures in the region ($T_{\rm d}$$\approx$100\,K) are
significantly lower than the inferred gas temperatures. Hence,
grain-gas collisions can also contribute to the gas cooling.

%V3
 \begin{table}[t]
\centering
\caption[]{Observed luminosities and cooling budget (assuming $d$=414\,pc)}
\begin{tabular}{ccccc}
\hline\hline
\multicolumn{1}{c}{Species} & 
\multicolumn{1}{c}{$L_{\rm obs}$ $(L_\odot)$} & 
\multicolumn{1}{c}{$\%$$^a$} & 
\multicolumn{1}{c}{$L_{\rm obs}$ $(L_\odot)$}  &
\multicolumn{1}{c}{$\%$$^b$} \\ 
     &  in Peak~1$^{c}$  & in Peak~1$^{c}$ & in map$^{c}$ & in map$^{c}$  \\ \hline
FIR $^{12}$CO     & 18.7          & $\sim$(22$\pm$5)\%   &  84  	       & $\sim$31\%  \\
FIR H$_2$O        & 7.0           & $\sim$(8$\pm$2)\%    &  32  	       & $\sim$12\% \\
FIR OH            & 3.5           & $\sim$(4$\pm$1)\%    &  13  	       & $\sim$5\% \\
FIR $[$\OI$]$     & 3.5           & $\sim$(4$\pm$1)\%    &  18  	       & $\sim$7\% \\
FIR $[$\CII$]$    & 0.3           & $\sim$(0$\pm$0)\%    &   7  	       & $\sim$3\% \\
FIR $^{13}$CO     & 0.3           & $\sim$(0$\pm$0)\%    &   2  		   & $\sim$1\% \\\hline
$^{12}$CO $v$=1-0 & $\sim$6.1$^d$ & $\sim$7\%    &  $\sim$14$^d$   & $\sim$5\% \\
H$_2$O  $v_2$=1-0 & $\sim$1.3$^d$ & $\sim$2\%    &  $\sim$3$^d$    & $\sim$1\% \\
MIR H$_2$         & $\sim$43$^e$  & $\sim$51\%   &   $\sim$102$^e$ & $\sim$37\% \\
Total     & $\sim$84              & 100\%        &   274           &    100\%  \\\hline
FIR dust  &  $\sim$4$\times$10$^3$   & - & $\sim$10$^5$ & \\\hline
\end{tabular}
\\{\vspace{0.2cm}}$^a$ In a $\sim$30$''$$\times$30$''$ aperture ($\sim$12500\,AU\,$\times$\,12500\,AU).\\
$^b$ In the entire $\sim$2$'$$\times$2$'$ map ($\sim$0.25\,pc\,$\times$\,0.25\,pc).\\
$^c$ Absolute calibration accuracy up to $\pm$30\%.\\
$^d$ Scaled from ISO/SWS observations \citep{Gon02}.\\
$^e$ Scaled from ISO/SWS observations \citep{Ros00}\vspace{0.5cm}.\\
\label{table:luminosities}
\end{table}

Assuming that our simple model  provides a good representation of the
average conditions toward Peak~1, we can also discuss the relative contribution of
H$_2$O, CO and OH cooling in the different temperature components. % to the far-IR line cooling. 
In our model we find that  CO in the \textit{hot} component ($T_{\rm k}$$\simeq$2500~K)
only contributes $\sim$1-2\%\, to the total CO rotational emission.
Indeed, the CO column density in the \textit{hot} component is $\sim$10$^3$ times smaller
than in the \textit{warm} component ($T_{\rm k}$$\simeq$500~K).
On the other hand, the H$_2$O emission from the \textit{hot}
component contributes with $\sim$40\% of the total H$_2$O rotational emission. 
In particular we predict a H$_2$O/CO$\simeq$1.3 abundance ratio and a H$_2$O/CO$\simeq$10 luminosity ratio  in the  \textit{hot} component alone.
Although these numbers are not exact and depend on the model assumptions, they suggest that
\textit{it is only in the \textit{hot} ($>$500\,K) postshock gas  where water vapor lines dominate the cooling over CO lines}.

Considering the rotational line cooling in both the \textit{hot} and \textit{warm} components together, 
the modelled H$_2$O/CO luminosity ratio is $\simeq$0.3.
For comparison, a H$_2$O/CO$\simeq$0.2  ratio was inferred
from the H$_2$O and  CO vibrational  bands \citep{Gon02}.
This  means that \textit{considering all components in the shock-excited gas toward Peak~1 together, 
CO, and not water vapor, is the most important gas coolant after H$_2$}.

\section{Shock(s) properties}\label{sec:shocks}

Fast dissociative shocks can destroy molecules and ionize atoms, whereas slower shocks  heat the gas
to high temperatures without destroying molecules 
\citep[e.g.,][for Orion-KL]{LB02}. 
Depending on the shock wave velocity (v$_{\rm s}$), pre-shock gas density and  magnetic
field strength and orientation %, and depending on the evolution of the shock structure 
it is common to distinguish between $J$-type (\textit{Jump}) and $C$-type (\textit{Continuous}) shocks. 
More complicated, ``mixed''  non-stationary situations ($CJ$-type)
may also exist \citep{Dra93,Wal05,Les04,Flo13}. 

Early steady-state planar shock models interpreted the available observations of Orion outflows in the frame
of a non dissociative $C$-type shock with preshock densities of several 10$^5$\,cm$^{-3}$, v$_{\rm s}$$\simeq$36-38\,km\,s$^{-1}$
and a  magnetic field perpendicular to the shock propagation  close to 1\,mG
\citep[see e.g.,][]{Dra82,Che82}. At these high shock velocities, 
vaporization of ice grain mantles is expected \citep{Dra83}.
In addition the gas reaches elevated postshock temperatures (well above $\sim$1000\,K).
Both effects lead to a drastic 
enhancement of the water vapor abundance.
These parameters of a single shock, however, should be taken as an average of the different shock-excited regions included
in the large $\sim$1$'$ beam of the early near- and far-IR observations \citep[][]{Gen89}.
Using a more detailed description of the gas-phase chemistry, molecular cooling and ion-neutral coupling,
new $C$-shock models confirmed those shock parameters and predicted that large abundances of 
H$_2$O, above 10$^{-4}$ with respect to H$_2$, will form in the  
postshock gas \citep[all the oxygen not incorporated in CO;][]{Kau96,Ber98}.
Alternative models try  to explain the H$_2$ rotational and vibrational line emission toward Peak~1 \citep{Ros00}
with a two-component $C$-shock with lower preshock density ($\sim$10$^4$\,cm$^{-3}$) and two shock
wave velocities,  v$_{\rm s}$=60 and 40\,km\,s$^{-1}$ \citep{LB02}. 
For the expected shock parameters in Orion, all these models predict that the H$_2$O line emission
will be the second more important gas coolant agent after H$_2$. They also predict high H$_2$O/CO abundance
ratios $>$1 in the postshock gas.
Our observations, however, show that H$_2$O lines (rotational and ro-vibrational) are only responsible of $\sim$10\%~of the
total luminosity emitted by H$_2$, CO, H$_2$O and O. In addition, the observed far-IR $L$(H$_2$O)/$L$(CO) luminosity ratio 
toward Peak~1 is $\simeq$0.4, at least an order of magnitud lower than the best $C$-shock model predictions for Orion.
Perhaps the most intriguing discrepancy  is the high observed far-IR
$L$(OH)/$L$(H$_2$O)$\simeq$0.5 luminosity ratio, either
suggesting that  water vapor formation is not so efficient or that 
an additional H$_2$O destruction mechanism (enhancing OH) exists.

\subsection{Low H$_2$O abundances: O depletion? Dissociation?}

Most early shock models in the literature assumed high initial abundances of either O$_2$ molecules or O atoms 
in the preshocked gas. Hence, for moderate shock velocities $\gtrsim$10-15\,km\,s$^{-1}$ 
(raising the postshock gas temperature to $T_{\rm k}$$\gtrsim$400\,K) most of the
available oxygen was converted into high abundances of water vapor through efficient gas-phase  neutral-neutral reactions 
\citep[e.g.,][and references therein]{Ber98}.
However, \textit{ISO}, \textit{SWAS}, \textit{ODIN}, and  most recently \textit{Herschel} observations suggest that there is far less H$_2$O 
\citep[][]{Ber02,Klo08,Cas10,Cas12} and O$_2$ \citep[][]{Gol00,Pag03,Gol11}
in quiescent dark clouds than would be expected if atomic oxygen were very abundant.
In fact, a wealth of observations  unambiguously show that most of the water in cold dark clouds is locked as ice mantles \cite[see][for review]{vD13}. 

Dark cloud observations also suggest that \textit{an undetermined fraction
of O atoms stick to grains, thus limiting the oxygen reservoir for water vapor formation}. %\citep[\textit{e.g.,}][]{Hol09}.
Indeed, recent investigations do show that atomic oxygen is being depleted even from diffuse clouds at a rate that cannot be accounted for
its presence in oxygen-rich dust grains \citep[][]{Jen09}. This implies that a unidentified reservoir of depleted oxygen is already present 
before the onset of freeze-out in cold dark clouds (before star-formation begins) thus reducing the availability of O atoms \citep[see also][]{Whi10}. 
Unfortunately, the adsorption energy of O atoms to ice is not fully constrained \citep[][]{Has93,Hol09}. 
Adsorption energies higher than the usually assumed value of $\sim$800\,K (similar to CO) would imply significant O atom depletion in molecular clouds
\citep[see][]{Mel12}. 

With fewer O atoms in the gas-phase than previously assumed and most of the H$_2$O locked as ice mantles, 
fast shocks ($\gtrsim$25\,km\,s$^{-1}$) are necessarily  needed to sputter the frozen-out H$_2$O \citep[][]{Dra95}
and produce high abundances of water vapor. 
Therefore, \textit{only fast shocks characterized by high postshock gas temperatures
($>$1000\,K) will show high gas-phase H$_2$O/CO$\gtrsim$1 abundance ratios} \citep[see also][]{Neu14}. 
Slower shocks, however,  heat the gas to a few hundred K but do not  sputter the ice mantles. 
\textit{Combined with a reduced initial abundance of gas-phase O atoms, slow shocks will produce regions with lower 
H$_2$O/CO abundance ratios ($\ll$1)}.

In the presence of strong UV radiation fields (see also Sect.~\ref{sect-UV}), water vapor is photodissociated and leads 
to high OH/H$_2$O$\gtrsim$1
abundance ratios 
\citep[see][for the Orion Bar PDR]{Goi11}. In addition, the endothermic reaction (a few
thousand K):
\begin{equation}
{\rm H_2O+H\rightarrow OH + H_2}
\end{equation}
can convert some H$_2$O into OH if the gas is hot enough, and if free H atoms are available 
(if a fraction of H$_2$ molecules are dissociated). 
These are conditions characteristic of, at least partially dissociative, $J$-shocks
\citep[][]{Neu89,Hol89}.
\HI\, observations  show that a fraction of H atoms do exit in Orion's outflows. 
In particular, the \HI\,  emission in the v$_{LSR}$=19 to 31\,km\,s$^{-1}$  range
follows  the orientation of the high-velocity outflow \citep{vDW13}.
Hence, it is plausible that \mbox{$J$-shock} spots with a non negligible H$_2$ dissociation fraction  exist.

A firmer evidence may come from the observed line surface brightness of species like OH or O.
Their emission lines are predicted to be weak in non-dissociative $C$-shocks and bright in dissociative  $J$-shocks.
As an example, the OH ground-state line at $\sim$119.4\,$\mu$m and the [\OI]\,63\,$\mu$m line
toward Peak~1 are very bright, $\sim$4$\times$10$^{-6}$\,W\,m$^{-2}$\,sr$^{-1}$
and $\sim$3$\times$10$^{-5}$\,W\,m$^{-2}$\,sr$^{-1}$ respectively. 
In the recent shock models of Flower \& Pineau des For{\^e}ts (2013), these intensities can only
be reproduced by relatively dense  preshock gas ($>$10$^5$\,cm$^{-3}$) subjected to $J$-type shocks. 
In this context, the decrease of the  observed H$_2$O 3$_{21}$-2$_{12}$/[\OI]\,145\,$\mu$m 
line intensity ratio (and increase  of the OH 84.6\,$\mu$m/[\OI]\,145\,$\mu$m ratio)  
toward Peak~1 (Figures~\ref{fig:observed_ratios}$e$ and \ref{fig:observed_ratios}$f$) are consistent with
H$_2$O dissociation in the shock.

\subsection{Mixed $C$- and $J$-type shocks} 

The strong H$_2$ line luminosities toward Peak~1 and their implied excitation temperatures
\citep[up to $\sim$3000\,K][]{Ros00} are not entirely compatible with dissociative $J$-shocks in which molecules are destroyed
by collisional dissociation.  Although the atomic gas is initially very hot (and UV radiation), by the time that molecules reform the gas has cooled to about 500\,K
\citep[e.g.,][]{Neu89,Hol89}. Therefore, $J$-type shocks are a plausible  source of
UV radiation in downstream molecular gas and may contribute to the
\textit{warm} CO component ($T_{\rm k}$$\simeq$500\,K) and that shows emission peaks both
toward Peak~1 and also $\sim$10$''$ NE of the Hot Core). These positions likely represent the apex of the
high-velocity and low-velocity outflows interacting with the ambient cloud.
Indeed, $\sim$10$''$ resolution  observations of the [\CI] $^3P_2$- $^3P_1$ and CO \mbox{$J$=7-6} lines with the
\textit{CSO}  suggest a $\sim$20$''$ shell structure of atomic carbon at the edge of the region where the
high-velocity CO emission is present \citep{Par05}. The [\CI] $^3P_2$- $^3P_1$ integrated line intensity
peaks  NE of the Hot Core and also toward Peak~1. The [\CI] emission was interpreted as CO dissociation in 
$J$-shocks. Interestingly, both the  OH/H$_2$O 84.6/75.4\,$\mu$m map in 
Fig.~\ref{fig:observed_ratios}$a$ as well as the CO 34-33/21-20  map
in Fig.~\ref{fig:lvg_ratios}$a$ suggest such a shell structure.

The strong H$_2$ lines in Orion are usually explained as the signature of 
non-dissociative shocks \citep[][and references therein]{LB02}.
The high kinetic temperatures that we infer in the \textit{hot} CO component suggest that 
\textit{significant fraction of hot H$_2$ and CO coexist}. 
However,  the broad range of  inferred H$_2$ excitation temperatures
are not entirely consistent with a single planar $C$-shock that, in  a first approximation, can be considered
as roughly isothermal  at the maximum shock temperature \citep[see][]{Ros00}.

$C$-type shocks have compression factors ($n/n_{\rm preshock}\lesssim$30) smaller than those in
$J$-shocks ($\lesssim$300). Hence, for the same preshock gas density, one could expect the densest shocked gas in
$J$-type shocks. Such shocks evolve faster and over smaller spatial scales \citep[e.g.,][]{Flo10}.
More importantly, given the short dynamical age of  the wide-angle outflow ($<$1000 yr) it is very likely
that the shock structures have not reached steady state. In this case, shocks in Orion's outflows likely have a range of
shock velocities and can show both $C$- and $J$-type attributes \citep{Chi98,Flo13}.

\subsection{UV-radiation diagnostics: irradiated shocks?}\label{sect-UV}

UV-irradiated $C$-shocks have not been previously considered to explain the far-IR emission from Orion but
 are an alternative solution to interpret the  high OH/H$_2$O abundance ratios
 while predicting very high postshock temperatures  and  bright H$_2$ lines.
C$^+$, with an ionization potential of 11.3\,eV, is the best probe of UV radiation able to dissociate
H$_2$O, CO and OH molecules. The [\CII]\,158$\mu$m and [\OI]\,145$\mu$m line surface brightness maps 
observed with \textit{Herschel}/PACS 
(Figures~\ref{fig:orion_maps_pacs}$b$ and \ref{fig:orion_maps_pacs}$c$) show a similar spatial distribution
that is almost orthogonal to that of the H$_{2}$ outflow and  to the high excitation CO, H$_2$O and OH lines.
As the [\CI] $^3P_2$- $^3P_1$ line \citep{Par05}, [\OI] peaks NE
of the Hot Core and also shows emission toward the Peak~1 region. 
They are consistent with molecular gas dissociation.

The spatial distribution
of [\OI]\,145/[\CII]\,158 line surface brightness ratio, however, is completely different and resembles the plateau,
reaching its maximum near Peak~1 (Figure~\ref{fig:oi_cii_obs_ratio}).
The [\OI]\,145/[\CII]\,158 intensity ratio peak value in this region is high, $\gtrsim$1.4, more than a factor 2 higher than the 
value observed in the Orion Bar PDR \citep{Ber12}.
Therefore, the regions shown in Fig.~\ref{fig:oi_cii_obs_ratio} where [\OI]\,145/[\CII]\,158$>1$ can
represent shocks where molecules are dissociated instead of PDR gas at the foreground \HII~region/OMC1  interfaces.

\begin{figure}[h]
\begin{center}
\vspace{0.3cm}
\includegraphics[angle=0,scale=.45]{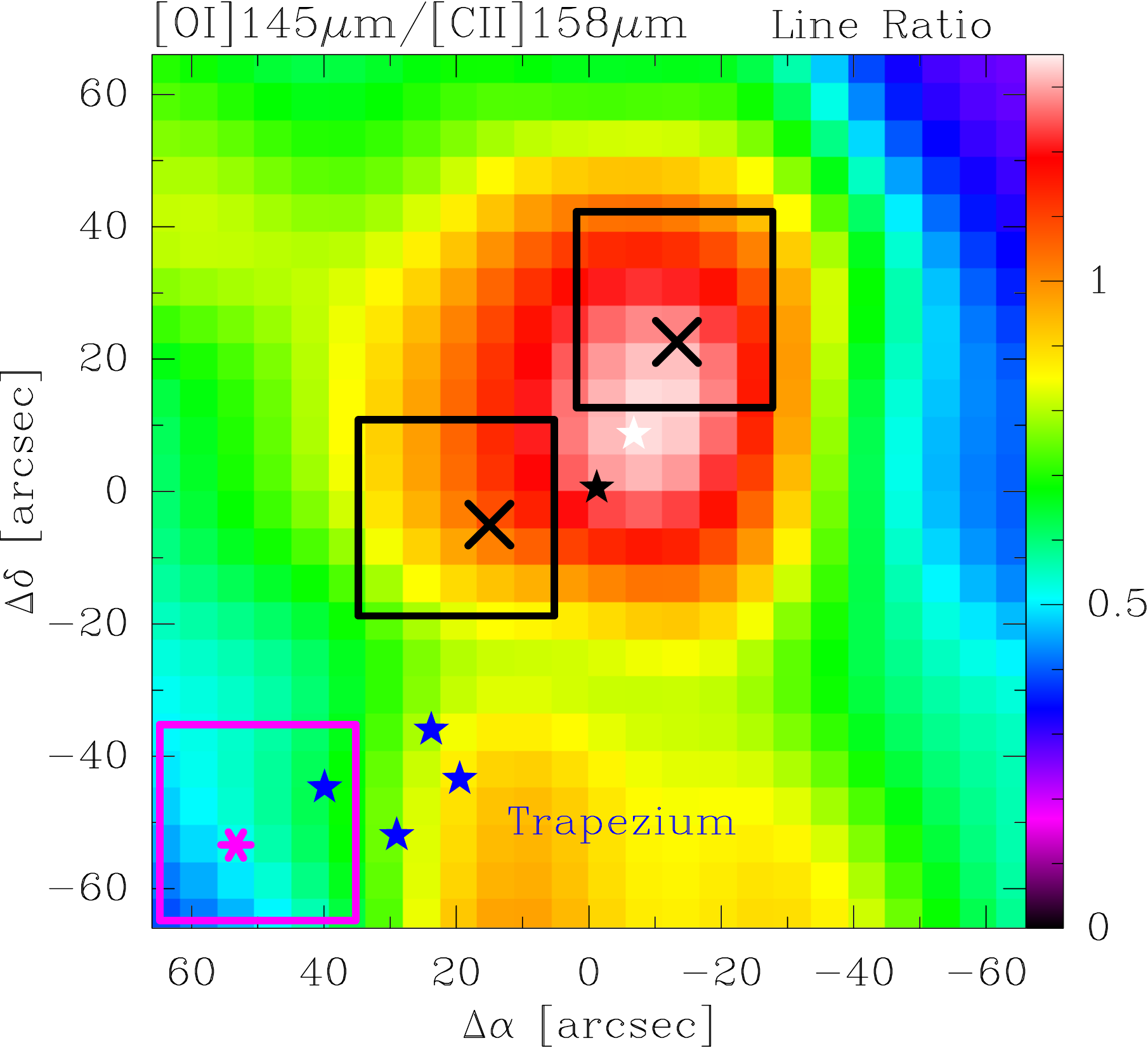} 
\caption{Map of the [\OI]\,145$\mu$m/[\CII]\,158$\mu$m surface brightness ratio toward Orion~BN/KL outflows obtained from \textit{Herschel}/PACS
observations.}
\label{fig:oi_cii_obs_ratio}
\end{center}
\end{figure}

Taking into account the FIR dust radiation field toward Peak~1   
($I_{\rm FIR}$$\simeq$0.034~W\,m$^{-2}$\,sr$^{-1}$)  and assuming that dust grains are heated by UV photons,
one can approximately constrain the UV field in the region \citep[e.g., 
by comparing with the grid of PDR models of][]{Kau99,Pou08}. 
We find that the
[\OI]\,145$\mu$m, [\CII]\,158$\mu$m and $I_{\rm FIR}$ surface brightnesses toward Peak~1 can be reproduced with two possible
combinations of gas density and UV field strength ($G_0$ in Habing's units); A lower density medium with $n$(H$_2$)$\simeq$10$^{4}$\,cm$^{-3}$
and $G_0$$\simeq$5$\times$10$^4$ (typical of the foreground PDR between the ionized Orion nebula and OMC1 and thus not related with the
star-forming cores) and a higher density medium with  $n$(H$_2$)$\simeq$10$^{6}$\,cm$^{-3}$ and  $G_0$$\lesssim$100.
This latter solution is consistent with the gas density in  Orion's outflows  and would imply that the postshock gas 
is illuminated by a moderate UV field, also produced \textit{in-situ} by fast $J$-shocks \citep[see also][]{Neu89b,Has91}.

\subsection{Is Orion-BN/KL a pathological star forming core?}

Early observations of Orion already suggested that there is not enough energy in the 
\HII~region or in the protostars embedded in the BN/KL region to provide the observed outflow(s) luminosity.
As a consequence, the bright H$_2$ and high-velocity CO outflow must have arised from a 
cataclysmic explosion \citep[e.g.,][]{Kwa76,Bec78}.
Today it is believed that the interaction of  sources BN, $I$ and $n$ 
caused an explosion responsible of the observed wide-angle H$_{2}$ outflow $\sim$500-1000\,yr ago
\citep[][]{Bal05,Gom05,Zap09,Nis12}. In fact, mergers in multiple massive stellar systems may be a
common feature in other high-mass star forming regions.

Far-IR observations toward more distant massive-star forming regions  can 
help to disentangle whether Orion~BN/KL is more a prototypical or a pathological example.
Karska et al. (2014) published PACS spectra toward 10
high-mass star forming sources 	\citep[from the \textit{WISH} Key Program sample,][]{vD11} 
and located at distances ranging from $\sim$1.5 to 5\,kpc (W33A, W3-IRS5, DR21(OH), NGC 6334-I, G34.26+0.15, etc.).
They analyzed the luminosities in the PACS central  spaxel ($\sim$10$''$) where the far-IR continuum peaks and 
most of the high-$J$ CO line flux arises ($\sim$10$''$$\simeq$0.25\,pc at a distance of 5\,kpc). 
In order to study similar spatial scales, one has to compare with
the 2$'$$\times$2$'$  area mapped by us in Orion~BN/KL ($\sim$0.25\,pc\,$\times$\,0.25\,pc)
(see the total luminosities in Table~\ref{table:luminosities}).
  
High-mass star-forming regions of the Milky Way's disk observed with \textit{Herschel} show a  CO 
rotational ladder  that can be approximately fitted with a 
single rotational temperature around $T_{\rm rot}$$\simeq$300$\pm$60\,K (line detections typically range from $J$=14 to 29). 
This component is  similar to the $\textit{warm}$ component inferred toward the 
Orion~Hot Core and H$_2$~Peak~1/2 regions from the CO $J$$<$30 lines.
The CO $J$=30-50 lines, however, are not detect in these more distant regions.
This is the range that we associate to a shock-excited \textit{hot} gas component 
toward Peak~1/2 (with $T_{\rm rot}$(CO)$\approx$1000\,K). Since these excited CO lines are detected in 
the shocks associated with much closer low-mass protostars \citep[e.g.,][]{Her12,Goi12,Man13} 
we suspect that their non-detection toward distant high-mass star-forming regions is  due to the lack of PACS sensitivity 
to detect weak lines at short wavelengths.

The far-IR CO lines in the sample of  Karska et al. (2014) show an mean absolute luminosity 
of $L_{\rm CO}$(PACS)=4$\pm$1\,$L_{\sun}$
if one excludes W51\,(G49.5-0.4) (that shows outstanding value of $L_{\rm CO}$(PACS)=25$\pm$11\,$L_{\sun}$
and $L_{\rm bol}$$\simeq$5$\times$10$^5$\,$L_{\sun}$).
Another extreme luminous massive star forming complex in the disk of the galaxy is W49N\,(G43.2-0.1), located
at a distance of 11.4\,kpc and radiating a luminosity of $L_{\rm bol}$$\simeq$10$^7$\,$L_{\sun}$ \citep{War90}.
Using unpublished \textit{Herschel}/PACS data\footnote{ObsIDs 1342207774 and 1342207775.} of W49N we extracted
$L_{\rm CO}$(PACS)=70$\pm$11\,$L_{\sun}$ toward the central spaxel.

In Orion~BN/KL, the far-IR CO luminosity in the $\sim$2$'$$\times$2$'$ map is $L_{\rm CO}$(PACS)=64$\pm$10\,$L_{\sun}$.
Together with W51 and W49N, these 3 high-mass star-forming regions  
show enhanced  CO luminosities compared to other massive regions in the galaxy disk.
Nevertheless, the normalized FIR CO luminosity ($L_{\rm CO}$(PACS)/$L_{\rm bol}$) in Orion~BN/KL is 
$\gtrsim$6$\times$10$^{-4}$, significantly higher than toward any of the above regions (more than an order of magnitude higher than 
their mean  5$\times$10$^{-5}$ or median 4$\times$10$^{-5}$ values).
Despite slightly different evolutive stages, this suggests that Orion BN/KL is a peculiar protostellar cluster,
characterized by violent shocks producing  higher $L_{\rm CO}$/$L_{\rm bol}$ luminosity ratios. 
This is likely the consequence of a recent explosive event (the merger).
The higher probability of protostellar encounters is likely related with the very high stellar density of Orion's core
 compared with other clusters in 
the Milky Way \citep[e.g.,][and references therein]{Riv13}.

Regarding water vapor, the observed $L$(H$_2$O)/$L$(CO)$\simeq$0.4 luminosity ratio toward Peak~1 shows that, overall, 
CO is more abundant and contributes
more to the gas cooling. However, a range of physical conditions and shock velocities likely exist in the shock-excited material
that are not resolved by \textit{Herschel}.
Our simple models suggest that H$_2$O is abundant and dominates the gas cooling over CO in the hottest molecular gas.
Neufeld et al. (2014) reached similar conclusions after a simultaneous fit of several H$_2$, H$_2$O and CO lines toward shocks in the protostellar
outflow  NGC\,2071 and in the supernova remnants W28 and 3C391. 
As in Orion~BN/KL outflows, only fast shocks  where the velocity is high enough (v$_{\rm s}$$>$25~km\,s$^{-1}$) to vaporize 
the H$_2$O ice grain mantles \cite{Dra95} are able to produce copious amounts of water vapor.

\section{Summary and conclusions}

We have presented $\sim$2$'$$\times$2$'$ spectral-maps of Orion~BN/KL outflows taken with \textit{Herschel}/PACS.
The much improved angular resolution compared to previous far-IR observations allowed us to spatially resolve, 
for the first time at these critical  wavelengths, the intrinsic emission from  H$_2$ Peaks~1/2 shocks
from that of the Hot Core and the ambient cloud.
The proximity of Orion  makes it a unique laboratory to study  the radiative and mechanical feedback  induced by 
massive stars and strong outflows at high spatial resolution. In particular, we obtained the following results:\\
$\bullet$~Peak~1/2 show a  rich  pure emisson spectrum, with more than 100 lines, 
most of them rotationally excited lines of $^{12}$CO (up to $J$=48-47 
with $E_{\rm u}$/$k$$=$6458\,K), H$_2$O, OH, $^{13}$CO, and HCN.
Around 10\% of the observed  line emission arises from  levels with 
$E_{\rm u}/k$$>$600\,K for H$_2$O and $E_{\rm u}/k$$>$2000\,K ($J$$>$27) for CO. 
Bright [\OI]\,63,145\,$\mu$m and fainter [\CII]\,158\,$\mu$m lines are also detected.\\
$\bullet$~More than half of the far-IR line luminosity toward Peak~1 is provided by CO lines.
H$_2$O ($\sim$20\%), OH ($\sim$10\%) and [\OI] ($\sim$10\%) lines have a less important contribution.
Including the mid-IR H$_2$ ro-vibrational lines and the CO and H$_2$O $v$=1-0  bands,
we estimate that $\sim$50\%~of the total line luminosity is emitted by H$_2$  followed by CO ($\sim$30\%) 
and H$_2$O ($\sim$10\%).\\ 
$\bullet$~The CO maps show an evolution of the spatial distribution of the high-$J$
CO lines that has not been reported before.
The brightest CO lines are those between $J$=15 and 20. They show an extended distribution with a half-power radius of $\sim$25$''$ around the Hot Core.
The $J$$>$30 lines however, peak toward Peak~1 and show a spatial distribution that resembles the wide-angle H$_2$ outflow
and the high-velocity (low$-J$) CO bullets observed from the ground. Better spectral resolution observations are clearly
needed to resolve the very excited far-IR CO high-velocity emission.\\
$\bullet$~Compared to other (sub)mm lines, the far-IR  H$_2$O and OH  lines show a different spatial distribution.
Most of them appear in absorption toward the strong far-IR continuum from the Hot Core and IRc sources. 
The same lines switch to emission toward Peak~1/2. 
Maps of different line surface brightness ratios suggest an increase of temperature and density toward Peak~1/2.\\
$\bullet$~The high-$J$ CO and OH lines are a factor $\approx$2 brighter toward Peak~1 whereas several
excited H$_2$O lines are $\lesssim$50$\%$  brighter toward Peak~2. The H$_2$O column and the gas density are likely
slightly higher towards Peak~2.\\
$\bullet$~The [\OI]\,145$\mu$m, [\CII]\,158$\mu$m and  $I_{\rm FIR}$ intensities toward Peak~1 can 
be reproduced with two possible combinations of $n$(H$_2$) and UV-radiation field. A lower density medium with
$n$(H$_2$)$\simeq$10$^{4}$\,cm$^{-3}$ and $G_0$$\simeq$5$\times$10$^4$ (typical of the foreground PDRs between the
 ionized Orion nebula and OMC1) and a higher density medium with  
 $n$(H$_2$)$\simeq$10$^{6}$\,cm$^{-3}$ (similar to the density in the outflows) and  $G_0$$\lesssim$100.
This latter combination would imply that the shocked gas in Orion~BN/KL is illuminated by a moderate UV-radiation field,
with significant \textit{in-situ} generated UV-photons   by fast $J$-type shocks.\\
$\bullet$~From a simple non-LTE model we conclude that most
 of the CO column density arises from $T_{\rm k}$$\sim$200-500\,K gas that we associate with low-velocity 
(v$_{\rm S}$$<$25~\,km\,s$^{-1}$) shocks that fail to sputter grain ice mantles and show a maximum gas-phase H$_2$O/CO$\leq$10$^{-2}$ abundance ratio. 
The very excited CO ($J$$>$35) and H$_2$O lines reveal a  hotter gas component ($T_{\rm k}$$\sim$2500\,K) from faster
shocks (with a smaller filling factor) that are able to sputter the frozen-out H$_2$O and lead to  high  H$_2$O/CO$\gtrsim$1 abundance ratios. 
The  moderate H$_2$O and high OH luminosities ($L$(H$_2$O)/$L$(CO)$\sim$0.4 and $L$(OH)/$L$(H$_2$O)$\sim$0.5)
cannot be reproduced by shock models   that assume high (undepleted) abundances of atomic oxygen in the preshocked gas and/or neglect the presence of
UV radiation in the postshock gas (triggering H$_2$O photodissociation).\\
$\bullet$~The  \textit{hot} shocked-gas  ($T_{\rm k}$$\approx$3500-1000\,K) inferred from
the CO $J$$\gtrsim$35 lines toward Orion H$_2$ Peaks  is not detected toward the Hot Core nor  toward 
more distant high-mass star-forming regions.
The total CO luminosity in these regions
(over spatial scales of $\sim$0.25\,pc; $\sim$10$''$
 at a typical distance of 5\,kpc) is usually more than an order of magnitude
smaller than $L$(CO)  over the entire Orion~BN/KL region (0.25\,pc$\simeq$2$'$).
Prominent  star-forming complexes like W51 or W49N 
also show comparably high CO luminosities, however, 
the normalized $L_{\rm CO}$(PACS)/$L_{\rm bol}$ luminosity
toward Orion~BN/KL is significantly higher. Orion thus seems more peculiar, 
probably because its strong outflows and associated shocks were caused by a recent explosive event.

\acknowledgments
We would like to thank the entire HEXOS GT-KP team for many useful and vivid discussions in the last years.
We thank Spanish MINECO for funding support under grants CSD2009-00038, AYA2009-07304 and AYA2012-32032.
J.R.G. was supported by a \textit{Ram\'on y Cajal} contract.

{\it Facilities:} \facility{Herschel Space Observatory}

\clearpage

%\end{document}

\appendix
\section{Line Fluxes toward H$_2$ Peak 1}

In the next  tables we present the observed line fluxes toward Orion~Peak~1 in the 
wavelength range $\sim$54-310\,$\mu$m obtained from \textit{Herschel}/PACS and SPIRE spectra 
in an aperture of $\sim$30$''$$\times$30$''$.

\begin{table}[h]
\centering
\footnotesize
\caption[]{CO line fluxes toward Orion Peak 1\label{tbl-fluxes-co}}
\begin{tabular}{ccccc}
\hline\hline
\multicolumn{1}{c}{Species} & 
\multicolumn{1}{c}{Transition} & 
\multicolumn{1}{c}{$\lambda$($\mu$m)} & 
\multicolumn{1}{c}{$E_{\rm u}$/$k$\,(K)}  &
\multicolumn{1}{c}{$F$~(W\,m$^{-2}$)$^a$} \\ 
$^{12}$CO & $J$=9-8 & 289.12 & 248.9 & 6.69E-14 \\
$^{12}$CO & $J$=10-9 & 260.24 & 304.2 & 9.18E-14 \\
$^{12}$CO & $J$=11-10 & 236.613 & 365.0 & 1.14E-13 \\
$^{12}$CO & $J$=12-11 & 216.927 & 431.3 & 1.41E-13 \\
$^{12}$CO & $J$=13-12 & 200.272 & 503.2 & 1.61E-13 \\
$^{12}$CO & $J$=14-13 & 185.999 & 580.5 & 1.84E-13 \\
$^{12}$CO & $J$=15-14 & 173.631 & 663.4 & 2.03E-13 \\
$^{12}$CO & $J$=16-15 & 162.812 & 751.8 & 2.21E-13 \\
$^{12}$CO & $J$=17-16 & 153.267 & 845.6 & 2.52E-13 \\
$^{12}$CO & $J$=18-17 & 144.784 & 945.0 & 2.43E-13 \\
$^{12}$CO & $J$=19-18 & 137.196 & 1049.9 & 2.27E-13 \\
$^{12}$CO & $J$=20-19 & 130.369 & 1160.3 & 2.13E-13 \\
$^{12}$CO & $J$=21-22 & 124.193 & 1276.1 & 2.04E-13 \\
$^{12}$CO & $J$=22-21 & 118.581 & 1397.4 & 1.80E-13 \\
$^{12}$CO & $J$=24-23 & 108.763 & 1656.6 & 1.43E-13 \\
$^{12}$CO & $J$=28-27 & 93.349 & 2240.4 & 6.15E-14 \\
$^{12}$CO & $J$=29-28 & 90.163 & 2399.9 & 5.73E-14 \\
$^{12}$CO & $J$=30-29 & 87.190 & 2565.0 & 4.57E-14 \\
$^{12}$CO & $J$=32-31 & 81.806 & 2911.3 & 3.96E-14 \\
$^{12}$CO & $J$=33-32 & 79.360 & 3092.6 & 3.37E-14 \\
$^{12}$CO & $J$=34-33 & 77.059 & 3279.3 & 3.32E-14 \\
$^{12}$CO & $J$=35-34 & 74.890 & 3471.4 & 2.94E-14 \\
$^{12}$CO & $J$=36-35 & 72.843 & 3669.0 & 2.06E-14 \\
$^{12}$CO & $J$=37-36 & 70.907 & 3871.9 & 1.68E-14 \\
$^{12}$CO & $J$=39-38 & 67.336 & 4293.9 & 7.18E-15 \\
$^{12}$CO & $J$=40-39 & 65.686 & 4512.9 & 5.84E-15 \\
$^{12}$CO & $J$=41-40 & 64.117 & 4737.3 & 5.13E-15 \\
$^{12}$CO & $J$=42-41 & 62.624 & 4967.1 & 4.51E-15 \\
$^{12}$CO & $J$=43-42 & 61.201 & 5202.2 & 3.57E-15 \\
$^{12}$CO & $J$=44-43 & 59.843 & 5442.6 & 4.15E-15 \\
$^{12}$CO & $J$=45-44 & 58.547 & 5688.4 & 3.24E-15 \\
$^{12}$CO & $J$=47-46 & 56.122 & 6195.8 & 2.48E-15 \\
$^{12}$CO & $J$=48-47 & 54.986 & 6457.5 & 2.08E-15 \\
\hline
$^{13}$CO & $J$=9-8 & 302.415 & 237.9 & 6.29E-15 \\
$^{13}$CO & $J$=10-9 & 272.205 & 290.8 & 1.90E-15 \\
$^{13}$CO & $J$=11-10 & 247.490 & 348.9 & 3.02E-15 \\
$^{13}$CO & $J$=12-11 & 226.898 & 412.4 & 4.86E-15 \\
$^{13}$CO & $J$=13-12 & 209.476 & 481.0 & 2.11E-15 \\
$^{13}$CO & $J$=14-13 & 194.546 & 555.0 & 2.50E-15 \\
$^{13}$CO & $J$=15-14 & 181.608 & 634.2 & 3.98E-15 \\
$^{13}$CO & $J$=16-15 & 170.290 & 718.7 & 3.61E-15 \\
$^{13}$CO & $J$=17-16 & 160.305 & 808.5 & 2.41E-15 \\
$^{13}$CO & $J$=18-17 & 151.431 & 903.5 & 2.33E-15 \\
$^{13}$CO & $J$=19-18 & 143.494 & 1003.8 & 2.15E-15 \\
\hline\hline
\end{tabular}
\\\vspace{0.2cm}$^a$In a $\sim$30$''$$\times$30$''$ aperture. Flux calibration accuracy up to $\sim$30$\%$.
\end{table}

\begin{table}[h]
\centering
\footnotesize
\caption[]{H$_2$O line fluxes toward Orion Peak 1\label{tbl-fluxes-h2o}}
\begin{tabular}{ccccc}
\hline\hline
\multicolumn{1}{c}{Species} & 
\multicolumn{1}{c}{Transition} & 
\multicolumn{1}{c}{$\lambda$($\mu$m)} & 
\multicolumn{1}{c}{$E_{\rm u}$/$k$\,(K)}  &
\multicolumn{1}{c}{$F$~(W\,m$^{-2}$)$^a$} \\ 
$o$-H$_{2}$O & $3_{12}-3_{03}$ & 273.193 & 215.2 & 1.33E-14 \\
$o$-H$_{2}$O & $3_{21}-3_{12}$ & 257.790 & 271 & 1.61E-14 \\
$o$-H$_{2}$O & $5_{23}-5_{14}$ & 212.526 & 608.2 & 3.57E-15 \\
$o$-H$_{2}$O & $2_{21}-2_{12}$ & 180.488 & 159.9 & 3.90E-14 \\
$o$-H$_{2}$O & $2_{12}-1_{01}$ & 179.527 & 80.1 & 8.72E-14 \\
$o$-H$_{2}$O & $3_{03}-2_{12}$ & 174.626 & 162.5 & 8.17E-14 \\
$o$-H$_{2}$O & $5_{32}-5_{23}$ & 160.504 & 697.9 & 5.23E-15 \\
$o$-H$_{2}$O & $5_{23}-4_{32}$ & 156.266 & 608.2 & blended\\
$o$-H$_{2}$O & $3_{30}-3_{21}$ & 136.494 & 376.4 & 3.14E-14 \\
$o$-H$_{2}$O & $5_{14}-5_{05}$ & 134.935 & 540.5 & 1.68E-14 \\
$o$-H$_{2}$O & $4_{23}-4_{14}$ & 132.407 & 397.9 & 4.45E-14 \\
$o$-H$_{2}$O & $4_{32}-4_{23}$ & 121.719 & 516.1 & 4.02E-14 \\
$o$-H$_{2}$O & $6_{16}-5_{05}$ & 82.030 & 609.3 & 5.95E-14 \\
$o$-H$_{2}$O & $4_{23}-3_{12}$ & 78.741 & 397.9 & 1.00E-13 \\
$o$-H$_{2}$O & $3_{21}-2_{12}$ & 75.380 & 271 & 1.12E-13 \\
$o$-H$_{2}$O & $3_{03}-2_{03}$ & 67.269 & 376.4 & 1.10E-14 \\
$o$-H$_{2}$O & $3_{30}-2_{21}$ & 66.437 & 376.4 & 3.60E-14 \\
$o$-H$_{2}$O & $7_{16}-6_{25}$ & 66.092 & 979 & 3.39E-15 \\
$o$-H$_{2}$O & $6_{25}-5_{14}$ & 65.165 & 761.3 & 1.10E-14 \\
$o$-H$_{2}$O & $8_{18}-7_{07}$ & 63.322 & 1036.5 & 4.28E-15 \\
$o$-H$_{2}$O & $4_{32}-3_{21}$ & 58.698 & 516.1 & 4.70E-14\\\hline
$p$-H$_{2}$O & $2_{02}-1_{11}$ & 303.459 & 100.8 & 2.06E-14 \\
$p$-H$_{2}$O & $1_{11}-0_{00}$ & 269.273 & 53.4 & 6.34E-15 \\
$p$-H$_{2}$O & $4_{22}-4_{13}$ & 248.247 & 454.4 & 4.52E-15 \\
$p$-H$_{2}$O & $2_{20}-2_{11}$ & 243.972 & 195.9 & 8.03E-15 \\
$p$-H$_{2}$O & $4_{13}-4_{04}$ & 187.110 & 396.4 & 1.10E-14 \\
$p$-H$_{2}$O & $3_{22}-3_{13}$ & 156.194 & 296.8 & blended\\
$p$-H$_{2}$O & $4_{31}-4_{22}$ & 146.919 & 552.3 & 8.96E-15 \\
$p$-H$_{2}$O & $4_{13}-3_{22}$ & 144.518 & 396.4 & 2.43E-14 \\
$p$-H$_{2}$O & $3_{13}-2_{02}$ & 138.527 & 204.7 & 9.14E-14 \\
$p$-H$_{2}$O & $3_{31}-3_{22}$ & 126.713 & 410.4 & 2.01E-14 \\
$p$-H$_{2}$O & $4_{04}-3_{13}$ & 125.353 & 319.5 & 6.24E-14 \\
%$p$-H$_{2}$O & $5_{33}-5_{24}$ & 113.944 & 725.1 & 1.25E-14 too high? \\
$p$-H$_{2}$O & $5_{24}-5_{15}$ & 111.626 & 598.9 & 8.91E-15 \\
$p$-H$_{2}$O & $3_{22}-2_{11}$ & 89.988 & 296.8 & 4.79E-14 \\
$p$-H$_{2}$O & $6_{06}-5_{15}$ & 83.283 & 642.7 & 1.84E-14 \\
$p$-H$_{2}$O & $7_{17}-6_{06}$ & 71.539 & 843.8 & 8.48E-15 \\
$p$-H$_{2}$O & $5_{24}-4_{13}$ & 71.066 & 598.9 & 1.08E-14 \\
$p$-H$_{2}$O & $3_{31}-2_{20}$ & 67.089 & 410.4 & 2.87E-14 \\
$p$-H$_{2}$O & $8_{08}-7_{17}$ & 63.457 & 1070.6 & 1.60E-15 \\
$p$-H$_{2}$O & $4_{31}-4_{04}$ & 61.808 & 552.3 & 2.02E-15 \\
$p$-H$_{2}$O & $8_{26}-7_{35}$ & 60.162 & 1414.2 & 4.03E-16\\
$p$-H$_{2}$O & $7_{26}-6_{15}$ & 59.986 & 1021.0 & 8.02E-16\\
$p$-H$_{2}$O & $4_{22}-3_{13}$ & 57.636 & 454.4 & 2.37E-14 \\
$p$-H$_{2}$O & $4_{31}-3_{22}$ & 56.324 & 552.3 & 1.53E-14 \\
\hline\hline
\end{tabular}
\\\vspace{0.2cm}$^a$In a $\sim$30$''$$\times$30$''$ aperture. Flux calibration accuracy up to $\sim$30$\%$.
\end{table}

\clearpage

\begin{table}[h]
\centering
\footnotesize
\caption[]{OH line fluxes toward Orion Peak 1\label{tbl-fluxes-oh}}
\begin{tabular}{ccccc}
\hline\hline
\multicolumn{1}{c}{Species} & 
\multicolumn{1}{c}{Transition} & 
\multicolumn{1}{c}{$\lambda$($\mu$m)} & 
\multicolumn{1}{c}{$E_{\rm u}$/$k$\,(K)}  &
\multicolumn{1}{c}{$F$~(W\,m$^{-2}$)$^a$} \\ 
OH & $^2\Pi_{1/2}~J=3/2^--1/2^+$ & 163.396 & 269.8 & 4.34E-014 \\
OH & $^2\Pi_{1/2}~J=3/2^+-1/2^-$ & 163.015 & 270.2 & 4.34E-014 \\
OH & $^2\Pi_{3/2}~J=5/2^+-3/2^-$ & 119.441 & 120.5 & 8.50E-014 \\
OH & $^2\Pi_{3/2}~J=5/2^--3/2^+$ & 119.234 & 120.8 & 8.50E-014 \\
OH & $^2\Pi_{3/2}~J=7/2^--5/2^+$ & 84.597 & 290.5 & 6.81E-014 \\
OH & $^2\Pi_{3/2}~J=7/2^+-5/2^-$ & 84.420 & 291.2 & blended \\
OH & $^2\Pi_{1/2}-^2\Pi_{3/2}~J=1/2^+-3/2^-$ & 79.179 & 181.7 & blended \\
OH & $^2\Pi_{1/2}-^2\Pi_{3/2}~J=1/2^--3/2^+$ & 79.116 & 181.9 & 1.14E-013 \\
OH & $^2\Pi_{1/2}~J=7/2^+-5/2^-$ & 71.215 & 617.9 & 1.38E-014\\
OH & $^2\Pi_{1/2}~J=7/2^--5/2^+$ & 71.171 & 617.7 & 1.38E-014\\
OH & $^2\Pi_{3/2}~J=9/2^+-7/2^-$ & 65.279 & 511.0 & 2.35E-014 \\
OH & $^2\Pi_{3/2}~J=9/2^--7/2^+$ & 65.132 & 512.1 & 2.36E-014 \\
OH & $^2\Pi_{1/2}~J=9/2^--7/2^+$ & 55.955 & 875.1 & 3.51E-015\\
\hline\hline
\end{tabular}
\\\vspace{0.2cm}$^a$In a $\sim$30$''$$\times$30$''$ aperture. Flux calibration accuracy up to $\sim$30$\%$.
\end{table}

\begin{table}[h]
\centering
\footnotesize
\caption[]{Atomic fine structure line fluxes toward Orion Peak 1\label{tbl-fluxes-atoms}}
\begin{tabular}{ccccc}
\hline\hline
\multicolumn{1}{c}{Species} & 
\multicolumn{1}{c}{Transition} & 
\multicolumn{1}{c}{$\lambda$($\mu$m)} & 
\multicolumn{1}{c}{$E_{\rm u}$/$k$\,(K)}  &
\multicolumn{1}{c}{$F$~(W\,m$^{-2}$)$^a$} \\ 
$[$\CII$]$  & $^2P_{3/2}-{^2P}_{1/2}$ & 157.741 & 91 & 6.30E-14\\
$[$\OI$]$   & $^3P_0-{^3P}_1$ & 145.525 & 327 & 8.68E-14 \\
$[$\OIII$]$ & $^3P_1-{^3P}_0$ & 88.356 & 163 & 3.51E-12 \\
$[$\OI $]$  & $^3P_1-{^3P}_2$ & 63.184 & 228 & 5.59E-13 \\
$[$\NIII$]$ & $^2P_{3/2}-{^2P}_{1/2}$ & 57.317 & 251 & 1.82E-14 \\
\hline\hline
\end{tabular}
\\\vspace{0.2cm}$^a$In a $\sim$30$''$$\times$30$''$ aperture. Flux calibration accuracy up to $\sim$30$\%$.
\end{table}

\end{document}